\ifdefined\pdfoutput\pdfoutput=1\fi %
\documentclass[manuscript,screen,nonacm]{acmart}
\AtBeginDocument{%
  }

\usepackage{amsmath}
\usepackage{booktabs}
\usepackage{graphicx}
\usepackage{subcaption}
\usepackage{arydshln} 
\usepackage{enumitem}

\begin{document}

\title{Blaming the Machine, Billing the Victim? How Mock Jurors Reassign Fault from Generative AI Systems and Respond to Liability Cliffs in Tort}

\author{Tin Trung Nguyen}
\email{tintn@umd.edu}
\affiliation{%
  \institution{University of Maryland}
  \city{College Park}
  \state{MD}
  \country{USA}
}

\author{Jiannan Xu}
\email{jiannan@umd.edu}
\affiliation{%
  \institution{University of Maryland}
  \city{College Park}
  \state{MD}
  \country{USA}
}

\author{Bao Kham Chau}
\email{bchau@umkc.edu}
\affiliation{%
  \institution{University of Missouri--Kansas City School of Law}
  \city{Kansas City}
  \state{MO}
  \country{USA}
}

\author{Hal Daum\'e III}
\email{hal3@umd.edu}
\affiliation{%
  \institution{University of Maryland}
  \city{College Park}
  \state{MD}
  \country{USA}
}

\author{Zubin Jelveh}
\email{zjelveh@umd.edu}
\affiliation{%
  \institution{University of Maryland}
  \city{College Park}
  \state{MD}
  \country{USA}
}

\renewcommand{\shortauthors}{Nguyen et al.}

\begin{abstract}
Generative AI complicates tort liability attribution when people blame a generative AI system that cannot itself bear legally relevant fault. We investigate where this blame goes when fault must be allocated among legal persons, and whether American discontinuous apportionment doctrines influence mock jurors’ factual findings of fault. Such doctrines include Contributory negligence, Joint and several liability, and their $50\%$ or $51\%$ variations, all of which have a fault-percentage threshold or a ``cliff'' where the victim-entitled compensation (or a defendant's liability) jumps to zero (or to all compensatory damages). Our pre-registered human subjects experiment (N=199 Prolific participants) shows that as many U.S. mock jurors ($36\%$) blame the AI system itself, the AI blame often ``leaks'' to the victim or to the AI user (in $27\%$ to $73\%$ of our AI-blaming records, depending on how leakage is calculated). Although mock jurors rarely change their blame assignment from one side of the liability cliff to the other, these participants do not acknowledge the legal doctrine or its monetary consequence when explaining their changes. Our empirical results illustrate how a narrow doctrinal exception in tort, carving out ``joint and several liability'' for those deploying or using Generative AI without reasonable care, might mitigate the impact of blame leakage on victims' deserved compensation without introducing significant jury nullification risks.\footnote{For the full survey, data, code, and other supplementary materials, please visit our Google Drive folder:
\url{https://drive.google.com/drive/folders/1zcweegD21f6LYxyWmqha0jitXwRZweO-?usp=sharing}
}

\end{abstract}



\maketitle

\section{Introduction}

In recent years, Generative AI (GenAI) systems have been rapidly adopted by the U.S. population. By June 2026, almost 50\% of U.S. adults have reported using GenAI chatbots \cite{gottfried2026americans}. Several databases, e.g., AIID \cite{paeth2024lessons} and AIAAIC \cite{aiaaic_repository}, as well as academic studies, have documented hundreds of AI-related incidents and demonstrated risks, such as 
deepfake and voice cloning scams \cite{deepfake, ftc2023scammers},
bad medical advice,~\cite{nyt2026chatgptseizure}
extraction of memorized training data~\cite{carlini2021extracting},
hate speech \cite{naztok}, 
fake news \cite{verma2023rise},
child sexual abuse materials \cite{USAttorney2023},
copyright disputes \cite{reed2023},
and chemical-weapon risks \cite{urbina2022dual}.
In 2026, agentic AI systems from OpenAI and Anthropic hacked into external organizations' cyber-infrastructure \cite{unpanel2026agents, anthropic2026incidents}.
These harms raise the pressing question of who should be held liable, both to deter future harm and to compensate victims.

A Generative AI system can produce harm through actions by several legal persons, e.g., the corporations developing and/or deploying the AI system, and the human users who prompt the AI system to produce certain outputs or perform certain actions, not to mention the actions taken by the AI system itself that go beyond the legal persons' expectations. Deciding whom to blame and how to apportion the blame across stakeholders is difficult, both morally and legally. The first complication in this blame attribution process is that people sometimes blame the AI system itself ~\cite{shank2018attributions,lima2023blaming}. However, under current tort doctrines, an AI system is not a legal person. It cannot be a defendant, bear legally relevant fault, or compensate a victim, so responsibility for its actions must be traced to a human or corporate actor (a legal person)~\cite {vladeck2014machines,selbst2020negligence}. Nevertheless, as AI systems do not appear as a party for fault attribution in real jury verdicts, we cannot directly observe whether that AI blame would have existed and to whom it would have gone: is it completely absorbed by the deployer, or would it leak to other parties in the case, such as the AI user (prompter) or the victim of harm, a phenomenon we define as ``blame leakage.''

In this paper, we use the noun ``blame'' not just in the moral sense, but interchangeably with ``fault'' or ``legally relevant fault,'' to mean mock jurors' factual findings and apportionment of faults, which, in a real trial, a judge would use to calculate liability for each defendant and compensation for the victim.

The next complication is how American apportionment doctrines, which translate jury-decided factual findings of fault into liability and compensation, may influence jurors' assigned faults in GenAI-related torts. To make these concrete, imagine a case in which the victim has suffered an indivisible injury worth 100,000~USD. Following the Restatement (Third) of Torts \cite{restatement2000apportionment}, we identify two major categories of such doctrines:
\begin{enumerate}[nolistsep,noitemsep]
\item \textbf{The victim's own fault.} While ``comparative negligence,'' a continuous doctrine with no cliff in compensation, simply deducts the victim's fault from their compensation (a victim at $1\%$ fault may recover 99,000~USD) ~\cite{li1975}, ``contributory negligence'' (CN), a discontinuous doctrine, bars the victim from any compensation if the jury finds the victim even partially at fault ~\cite{coleman2013}.
\item \textbf{Liability between multiple defendants.} While ``proportionate liability'' (also known as ``several liability''), a continuous doctrine with no cliff in either defendant's liability, holds each defendant liable for only their own portion of fault \cite{brown1978keill}, ``joint and several liability'', a discontinuous doctrine, allows the victim to recover all entitled compensatory damages from any one of the defendants, as long as that defendant is partially at fault ~\cite{NorfolkAyers2003}. In the example above, suppose Defendant A is at $1\%$ fault but has much more money than Defendant~B ($98\%$ fault). Under ``joint and several liability'' (JSL), after deducting 1,000~USD from the 100,000~USD loss due to their $1\%$ fault, the victim may opt to recover their entitled 99,000~USD damages from Defendant A alone.
\end{enumerate}
We characterize both ``contributory negligence'' and ``joint and several liability'' as \textbf{discontinuous doctrines} because when framed in a percentage-based format, their liability cliff effects trigger at any fault above a $0\%$ threshold. Popular compromises between those continuous and discontinuous doctrines within each category often choose a different threshold to trigger the respective liability cliff, most commonly at $50\%$ or $51\%$ fault, as recently surveyed across U.S. states by Matthiesen, Wickert \& Lehrer, S.C. \cite{mwl2022jointseveral, mwl2026comparativefault}. In this study, we investigate contributory negligence (CN), joint and several liability (JSL), and their half-way variants: modified comparative negligence with a $50\%$ bar (MCN50), where victim at half or more fault gets nothing \cite{mcintyre1992}, and modified joint and several liability with a $51\%$ threshold (MJSL51), where a majority-fault defendant might be liable for all compensatory damages \cite{texas33013}. We use comparative negligence and proportionate liability as a baseline to study how discontinuous doctrines affect mock jurors when finalizing blame.

\begin{figure}[tp]
    \centering
    \begin{subfigure}{0.65\textwidth}
        \centering
        \includegraphics[width=\textwidth]{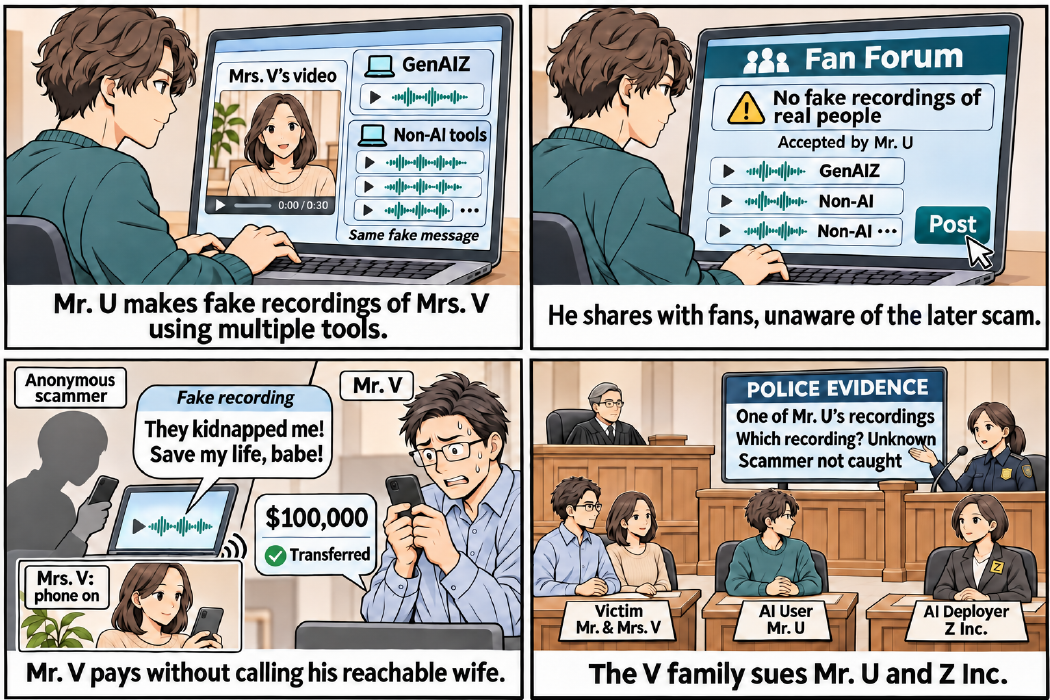}
        \caption{Scenario~1: Fake Kidnapping, designed to test if mock jurors tweak assigned faults to activate ``Joint and several liability'' or to nullify ``Contributory negligence'' }
        \label{fig:scenario1}
    \end{subfigure}

    \vspace{1.5em}

    \begin{subfigure}{0.65\textwidth}
        \centering
        \includegraphics[width=\textwidth]{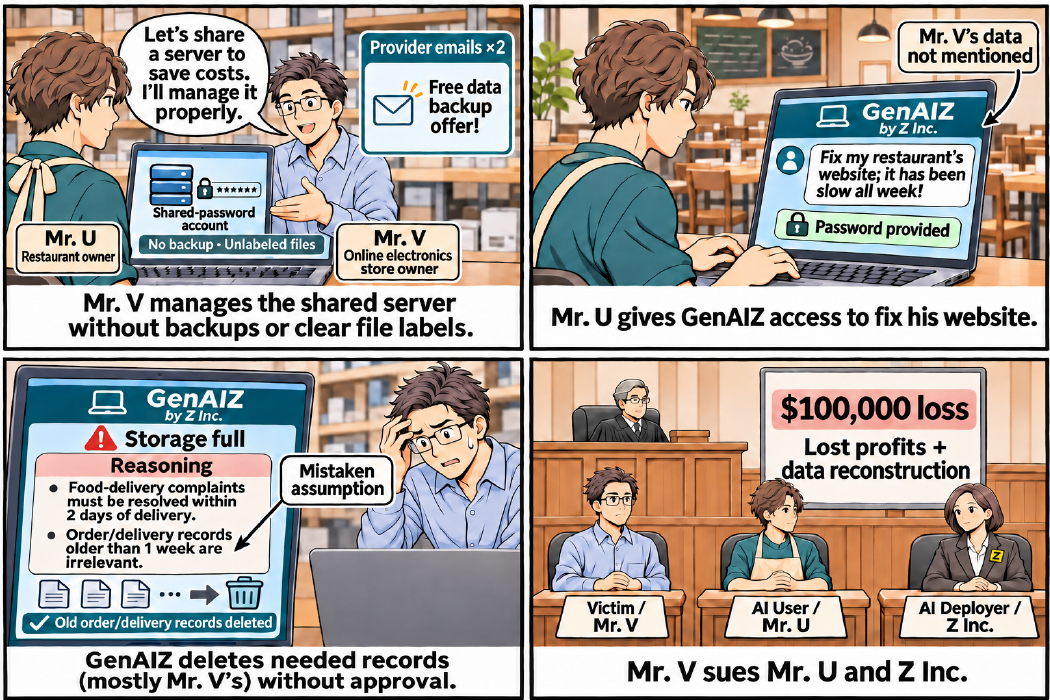}
        \caption{Scenario~4: Data Loss, designed to test if mock jurors tweak assigned faults to activate ``Modified joint and several liability, $51\%$ threshold'' or to nullify ``Modified comparative negligence, $50\%$ bar''}
        \label{fig:scenario4}
    \end{subfigure}
    \caption{Two of our four GenAI-involved harm scenarios, each illustrated in four panels (read left to right, then top to bottom). These comics were generated by GPT-6 Astra for illustration here only. Our participants were shown the text descriptions in \autoref{app:scenarios}.}
    \label{fig:scenarios}
\end{figure}

Our complete study is based on a preregistered, three-phase experiment with 199 U.S. participants from Prolific. At the start, each participant was randomly assigned to judge two of four GenAI-involved scenarios of unintentional torts: fake kidnapping, weight-loss plan, leaked manuscript, and data loss (we illustrate the first and the last in \autoref{fig:scenarios}, and show all four scenarios' full-text descriptions in \autoref{app:scenarios}). Each phase proceeded as follows:

\begin{enumerate}[nolistsep,noitemsep]
    \item[P1:] Mock jurors (Prolific participants) answered binary blame-factor questions and allocated fault among the victim, deployer, user, and AI system across two scenarios.
    \item[P2]: Mock jurors reallocated fault after being told that the system was not a legal person and could not bear fault. 
    \item[P3]: Mock jurors were instructed on either two baseline doctrines or one discontinuous doctrine paired with the remaining baseline doctrine, then repeated the three-party fault allocation task similar to Phase~2, but with a live calculator showing how their assigned fault percentages translate into liability and compensation.
\end{enumerate}

Our main findings are: First, many mock jurors (one-third of our participants) attributed non-zero blame to the AI system in Phase 1, and that blame often leaked to the victim and the AI user in Phase 2. Second, disclosing a discontinuous apportionment doctrine rarely influences mock jurors' factual findings of fault, but when it does, mock jurors tend to justify the tweak in their verdict by a new argument or factual assumption unrelated to the doctrine.

\section{Background and Related Work}
\label{sec:background}

\subsection{Blame and Mind Perception}
\label{sec:bg-blame}

\begin{figure}[t]
    \centering
    \includegraphics[width=0.6\textwidth]{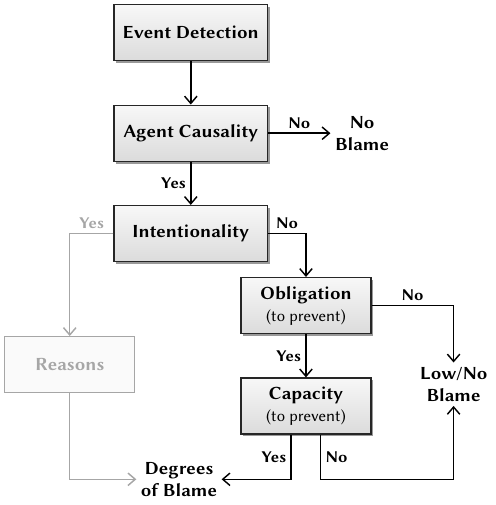}
    \caption{The Path Model of Blame by \citet{malle2014theory}, reproduced with the intentional path blurred out as our study concerns unintentional torts.}
    \label{fig:malle_path_model}
\end{figure}

The Path Model of Blame by \citet{malle2014theory} considers causality and intentionality, linking blame for unintentional outcomes to obligations and capacities to prevent them. We translate these blame factors, as reproduced in \autoref{fig:malle_path_model}, into binary questions to estimate the degree of legally relevant fault across different paths taken by the three legal persons in each scenario.

Research on mind perception distinguishes agency from experience~\cite{gray2007dimensions}. These perceptions vary across AI systems~\cite{ladak2025robots}, and people attribute wrongness, blame, and even culpable mental states to machines. When vignettes based on publicized AI failures described a moral violation, participants attributed more wrongness to the AI but not to the organization, programmers, or users behind it~\cite{shank2018attributions}. In another experiment, blame and attributed mental states were relatively similar for a human, a group agent, and an AI-driven robot~\cite{stuart2021guilty}. The direction of a human--machine difference depends on the comparison: \citet{awad2020drivers} found reduced blame for the automated member of a driving team when both drivers erred, whereas \citet{franklin2021blaming} found greater blame for automated drivers across their tested situations.

Moral blame also differs from financial responsibility. \citet{lima2023who} found that respondents assigned responsibility for machine-caused harms to both developers and users, with higher expected payments from developers. They distinguish attitudes toward blaming systems from judgments about their designers and users~\cite{lima2023blaming}. Our contribution is to compare allocations before and after removing the AI as an available fault recipient. This connects to concerns about gaps in responsibility~\cite{matthias2004responsibility}, institutional accountability~\cite{cooper2022accountability}, and excessive responsibility placed on the nearest human operator as defined in the ``moral crumple zone'' by \citet{elish2019moral}, with which our ``blame leakage'' concept resonates.

\subsection{Liability for AI-Related Harms}
\label{sec:bg-liability}

The AI liability scholarship examines negligence standards for automated systems and their users~\cite{abbott2018reasonable,selbst2020negligence}, product liability approaches~\cite{vladeck2014machines,chagal2019am}, analogies to employer responsibility~\cite{lior2020ai}, and the incentives created by assigning liability along the AI supply chain~\cite{buiten2023law} without automatically transferring an AI system's assigned blame to its deployer.

In the US, Colorado's 2026 law replacing its AI Act addresses fault allocation directly. In actions under state anti-discrimination law arising from consequential decisions materially influenced by covered automated decision-making technology, fault is allocated among developers and deployers according to their relative fault~\cite[\S~6-1-1707]{coloradoSB26189}.

\subsection{Verdict Consequences and Jury Nullification of Legal Doctrines}
\label{sec:bg-juries}

Best and Donohue~\cite{best2012jury} found that juries in modified comparative-negligence jurisdictions with a $51\%$ bar were less likely than those in pure comparative-negligence jurisdictions to assign victims above 50\% fault. However, this comparison is observational, not causal. Experimental work also shows that instructions about verdict consequences can affect decisions, although Diamond and Casper's study concerned treble damages in antitrust litigation, not fault thresholds~\cite{diamond1992blindfolding}. \citet{hans2000business} found no support for the simple deep-pockets account of corporate liability judgments. The studied jurors held corporations to a higher standard of conduct than individuals and were more suspicious of injured individuals.

\section{Design and Set-up of Human Subjects Experiment: Attributing Legally Relevant Faults as Mock Jurors}

\subsection{Research Questions}
We have pre-registered the following research questions for our human subjects experiment on mock jurors' attribution of fault in tort scenarios related to Generative AI.

\textbf{RQ0}: Do the blame factors from the Path Model of Blame by \citet{malle2014theory} reliably predict the relative ordering of legally relevant faults among the three main legal persons in GenAI-related torts (the victim, user, and deployer)?\footnote{In the real world, the deployer of a Generative AI system might not be the same as the developer of the underlying large language model (for example, Microsoft deploying Bing, a chatbot based on OpenAI's GPT-4 model \cite{battle2023microsoft}). In those cases, since the deployer and the developer are likely corporations, they are well-positioned to negotiate in their contracts (before harm) or during settlement (after harm) on how they will share the risks and liability when the AI system malfunctions, so the matter of blame and liability attribution between them falls more in the realm of contract law than tort law.}

\textbf{RQ1}: Do people often assign non-zero legally relevant faults to the AI System itself (Phase~1)?

\textbf{RQ1.1}: If Yes to RQ1, does this blame-the-AI pattern correlate with people’s tendency to anthropomorphize daily generative AI systems (as measured by four end-of-study questions along two dimensions: agency and experience)? 

\textbf{RQ1.2}: If Yes to RQ1, what are the main patterns of reassigning the AI system’s fault after people learn that they cannot assign any legally relevant fault to the AI system?

\textbf{RQ2}: Does disclosing one of the discontinuous apportionment doctrines, namely joint and several liability (JSL); contributory negligence (CN); modified joint and several liability, 51\% threshold (MJSL51); modified comparative negligence, 50\% bar (MCN50), change mock jurors’ factual findings of fault in GenAI-related torts?

\subsection{Experimental Design}
\begin{figure}[t]
    \centering
    \includegraphics[width=1.0\textwidth]{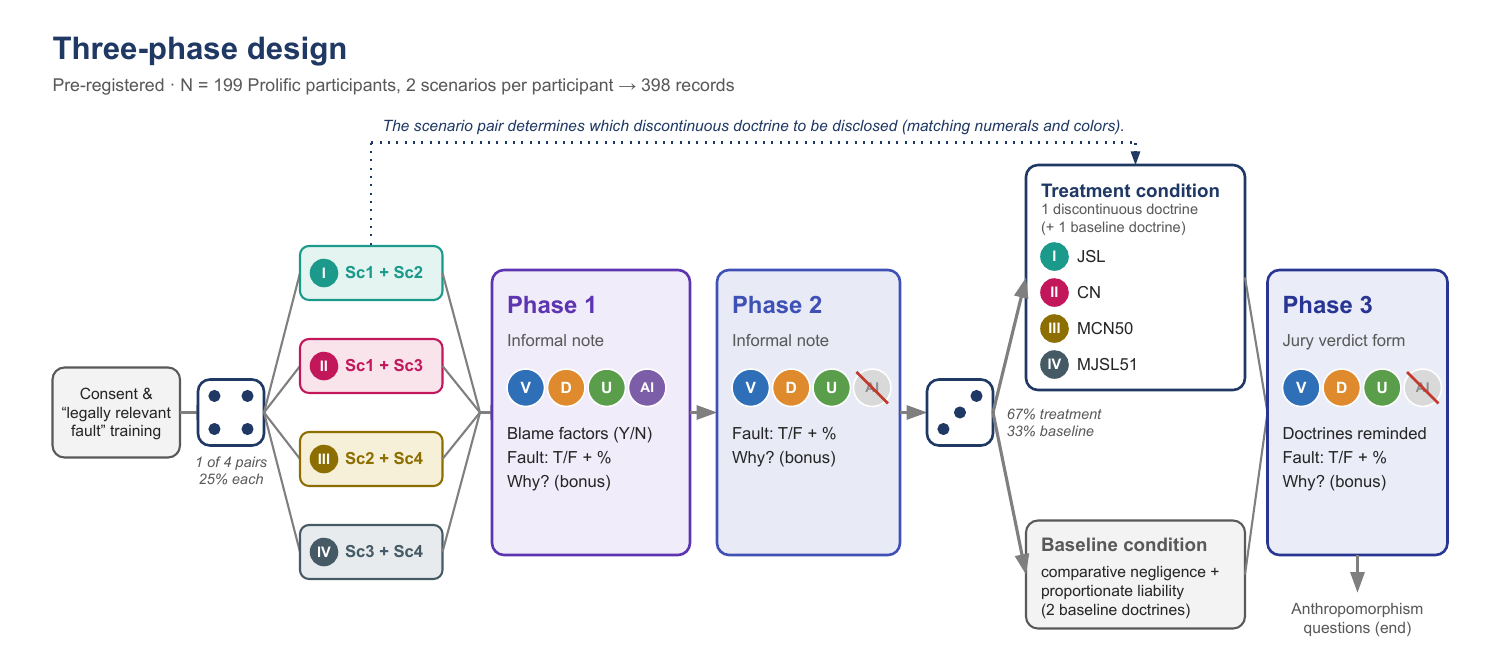}
    \caption{Our three-phase experimental design. Participants do both assigned scenarios in one phase before moving to the next. Fault attribution parties: V = victim, D = AI deployer, U = AI user, AI = AI system. Scenarios: Sc1 (fake kidnapping), Sc2 (weight-loss plan), Sc3 (leaked manuscript), Sc4 (data loss). Doctrines in treatment conditions: I. joint and several liability (JSL), II. contributory negligence (CN), III. modified comparative negligence, 50\% bar (MCN50), IV. modified joint and several liability, 51\% threshold (MJSL51).}
    \label{fig:design}
\end{figure}

We aim to address the following research questions (RQs) in our human subjects experiment, as designed in \autoref{fig:design}.

Before entering the main study, participants must complete a skill-screening and concept-training phase to understand what they will later allocate: ``legally relevant fault''. In American tort law, negligence generally requires a duty, breach, factual causation, a legally sufficient connection to the injury, and damages. Our pre-Phase~1 training operationalizes three of these: breach as the lack-of-care test, factual causation as the but-for test, and the connection to the injury as the foreseeability test \cite{restatement2010physical}. 
Our but-for and foreseeability tests also form the basis of the ``proximate cause'' standard \cite{doe1995boysclubs}, which is often asked in jury verdict forms \cite{block2009mora} and may in turn decide if a victim/defendant's fault triggers JSL/CN.

Randomization happens twice in the main study to address RQ2. For RQ0 and RQ1, we use observational methods. 

The first randomization happens right before Phase~1, when each participant is randomly assigned to one hidden “scenario tag” (one stratum, or pair of scenarios) out of four scenario tags, namely “12”,”13”, “24”, and “34”, with equal probabilities (25\%). This scenario tag decides which two (out of four) scenarios they will work on throughout all three phases of the study. Each scenario is designed to imply a baseline fault distribution that is close to the threshold of two discontinuous apportionment doctrines (one on the victim side and the other between the two defendants), thereby conducive to testing the jury nullification effect of those two doctrines. Two scenarios within each scenario tag are therefore conducive to testing one shared doctrine. 
In Phase~1, each participant will work on two scenarios. In each scenario, they will first read one paragraph of scenario description, followed by an attention check: they will answer Yes or No to 9-12 blame factor questions (“Victim Causality = ?” or “Did the Victim clearly have some causal contribution to the harm?”) based solely on a one-line description per blame factor (e.g., “Mrs. V’s TikTok videos enabled voice cloning. Mr. V’s failure to verify made the scam succeed.”). Each participant must answer at least 60\% of the questions correctly within an attention check (above the 50\% random guess baseline) to pass that attention check. We exclude data from any participants who fail both attention checks (both are in Phase~1 only). After completing an attention check, the participant will answer a binary True/False question on whether each of the following four parties has any non-zero legally relevant fault in the scenario: the victim, the AI deployer, the AI user, and the AI system. Then,  the participant will answer a numerical (percentage) fault assignment question for those four parties.

In the transition from Phase~1 to Phase~2, we only disclose that the AI system itself is not a legal person and thus will be excluded from fault attribution in Phase~2, where participants can attribute legally relevant faults to only the victim, AI deployer, and AI user. The scenarios and fault question format remain the same between those two phases.

The second randomization is in the transition from Phase~2 to Phase~3, when each participant is randomly assigned to either a baseline condition (one-third probability) or a treatment condition (two-thirds probability). Since each scenario is present in two scenario tags, these probabilities ensure data balance: for each scenario (across two tags), we will have roughly the same number of participants working on that scenario in the baseline condition compared to each treatment condition. In the baseline condition, the pre-Phase~3 transition will teach each participant two continuous apportionment doctrines (comparative negligence and proportionate liability). In contrast, in the treatment condition (two-thirds probability), the pre-Phase~3 transition will teach each participant one discontinuous apportionment doctrine that corresponds to their previously assigned scenario tag (for example, we teach “joint and several liability” for those with scenario tag “12” and assigned to the treatment condition in Phase~3) and one continuous apportionment doctrine that handles the victim side or the between-two-defendants relationship, whichever is not touched upon by the discontinuous doctrine yet (for example, comparative negligence). As participants answer the percentage allocation questions for legally relevant faults in Phase~3, they will also see in real time the monetary consequences of the two doctrines we have disclosed to them, as we have coded up the logic of each set of two doctrines into an interactive dashboard that calculates based on participant-entered percentages and displays the “Maximum Deployer Liability” (how much the victim may demand the deployer to pay), “Maximum User Liability” (how much the victim may demand the user to pay), and “Maximum Total Compensation” (how much the victim may recover from the deployer and the user combined when both parties have money to pay). When the User has no money, the Victim can recover only up to the Maximum Deployer Liability. However, we do not disclose any party’s ability to pay, because it is generally irrelevant to factual findings of fault and compensatory damages (Fed. R. Evid. 401–402), and courts may exclude such evidence when its probative value is substantially outweighed by the risk of unfair prejudice (Fed. R. Evid. 403) \cite{FRE}.

At the end, we ask people to think about actual GenAI systems they use daily and answer four sliding-scale questions on two dimensions of AI mind perception (agency and experience), as originally developed by \citet{gray2007dimensions}
and adapted to the AI context by \citet{ladak2025robots}. We use these ratings as a measure for AI anthropomorphism.

\begin{figure}[t]
    \centering
    \begin{subfigure}[b]{0.49\textwidth}
        \centering
        \includegraphics[width=\textwidth]{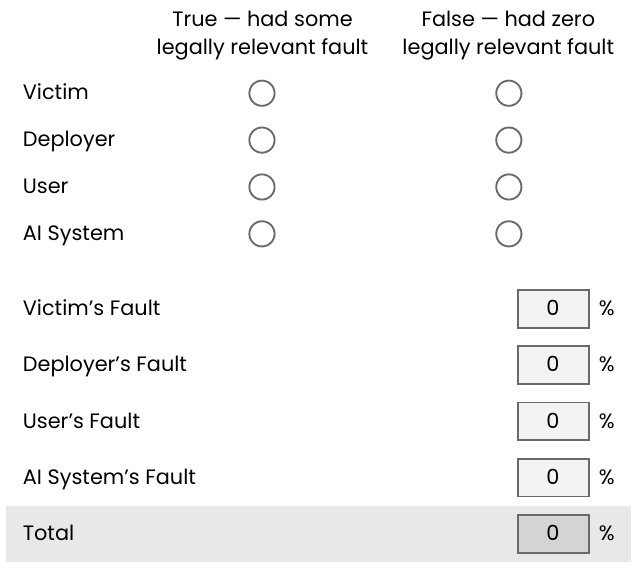}
        \caption{Phase~1: participants first judge whether each of the four parties (victim, AI deployer, AI user, and AI system) had any legally relevant faults, then apportion percentages for such faults.}
        \label{fig:interface_p1}
    \end{subfigure}
    \hfill
    \begin{subfigure}[b]{0.49\textwidth}
        \centering
        \includegraphics[width=\textwidth]{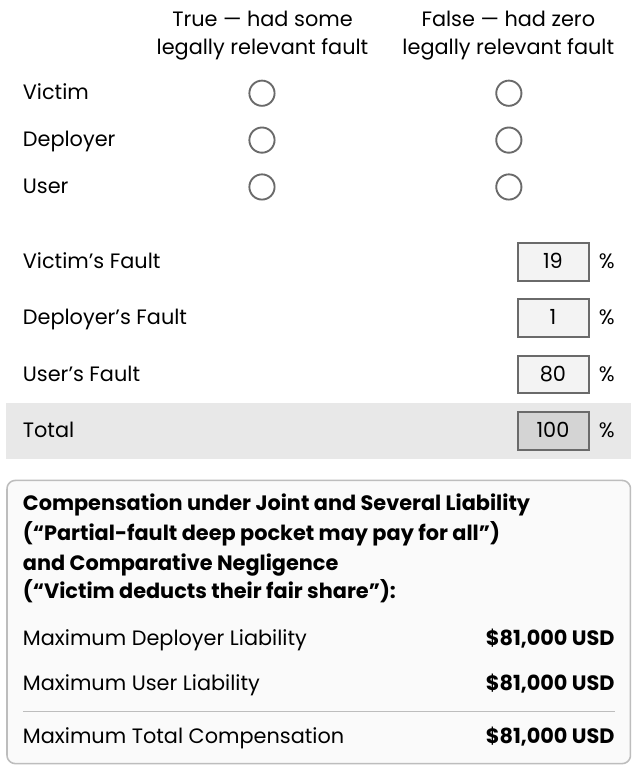}
        \caption{Phase~3 (example inputs): with the AI system excluded since Phase~2, participants attribute legally relevant faults among the three remaining parties while being informed in real time of the doctrine-driven money implications of their factual findings.}
        \label{fig:interface_p3}
    \end{subfigure}
    \Description{Two screenshots of the survey interface. Left: Phase~1 form with True/False fault questions and fault-percentage boxes for the victim, deployer, user, and AI system. Right: Phase~3 form with the same questions for the victim, deployer, and user, plus a panel showing each defendant's maximum liability and the maximum total compensation.}
    \caption{Qualtrics experiment interface of the main fault attribution questions in (a) Phase~1 and (b) Phase~3 of our study. Phase~2 looks like Phase~1 but without the ``AI System'' in either question.}
    \label{fig:interface}
\end{figure}

We show the format of our fault attribution questions in \autoref{fig:interface}.
We include our experimental setup with Prolific logistics in \autoref{app:setup} and the main content of our Qualtrics survey from \autoref{app:training} to \autoref{app:sliders}.

Our human subjects experiment was IRB-approved and pre-registered via the Open Science Framework (OSF).\footnote{Pre-registration proof: \url{https://osf.io/87faz/overview?view_only=a4ffbac17469482dae8ba2ced40e78c0}}
\section{To RQ0: The Path Model of Blame Can Reliably Compare Legally Relevant Faults Across Different Paths}

\label{sec:rq0}

Before we test our main research questions on blame leakage and mock jurors' responses to discontinuous apportionment doctrines, we first address RQ0 and test whether the blame factors from the Path Model of Blame (Causality, Intent, Obligation to prevent, Capacity to prevent; we exclude ``Reasons'' since our scenarios only deal with unintentional torts)~\cite{malle2014theory} can reliably predict how, in each scenario, the faults of the three legal persons compare to one another.

Based on participants' Yes/No blame factor answers, we test whether the following hierarchy held for our notion of ``legally relevant fault'' across the three legal persons (we disregard the AI system's fault here): rank 0 for non-causation; ranks 1--3 for unintentional causation with neither (rank 1), one (rank 2), or both of an obligation and capacity to prevent harm (rank 3); rank 4 for intentional causation. Our preregistered rule required 1/ higher-ranked parties to receive at least as much fault, 2/ equal-ranked parties' shares to differ by at most 10 percentage points, and 3/ non-causal parties to receive at most 10\% fault. It held in 44.7\% of Phase~1 records and 45.5\% of Phase~2 records, well above the 8.5\% benchmark of allocations drawn uniformly from the 231 three-party allocations on a five-point grid. Participants' own factor answers also predicted their allocations better than other participants' answers about the same scenarios as a second benchmark (by 7.6 and 8.6 points; permutation $p < .001$).\footnote{
We analyzed 383 and 385 eligible records, respectively, from 198 participants. A deployer judged causal in Scenarios~1--2, which asked only about its causality, has no rank. Two Phase~1 records gave the AI system 100\%. Phase~1 shares were rescaled to exclude the AI system. Hierarchy rates are averaged within participants. Our permutation benchmark rescores each participant's allocations against another participant's factor profiles from the same scenario pair (1,000 permutations; 185 participants with eligible records in both scenarios and phases). Confidence intervals in this section and in \autoref{sec:ai-blame-leak} are percentile intervals from 3,000 participant bootstrap resamples within scenario-pair strata. For the $\kappa$ difference, both terms use the same resamples.} 
 
Removing the equal-rank requirement increases agreement to 79.3\% in Phase~1 and 78.3\% in Phase~2: paired gains of 34.6 percentage points (95\% CI [30.2, 39.1]) and 32.8 points [28.4, 37.3]. 
Relaxing the rule necessarily raises or preserves agreement, so the gain alone cannot establish better performance on ordering than on equality. Thus, we compare these predictions among pairs judged causal and unintentional. In Phase~2, higher-ranked parties received at least as much fault in 84.0\% of 381 pairs; parties with identical factor profiles received shares within 10 points in 37.3\% of 287 pairs. Randomly reassigning each record's three shares to its parties yielded reference agreement of 57.6\% and 26.9\%, respectively. Using $\kappa=(p_{\mathrm{observed}}-p_{\mathrm{reference}})/(1-p_{\mathrm{reference}})$, agreement beyond this reference was 0.62 for ordering and 0.14 for approximate equality, a difference of 0.48 [0.38, 0.58]. The difference is still positive under strict ordering (0.40 [0.31, 0.49]), under a uniform-allocation reference (0.47 [0.36, 0.58]), and under both together (0.30 [0.18, 0.42]).
 
Our results indicate that a binary operationalization of the Path Model of Blame \cite{malle2014theory} could reliably compare the ``legally relevant fault'' of two legal persons positioned on two different blame-factor-based paths, but only weakly supports the prediction that two legal persons on the same path receive similar amounts of ``legally relevant fault''.
 
\section{To RQ1s: Blame Leaks from the AI System to the Victim or the User, even after Proportional Rescaling}

\label{sec:ai-blame-leak}
Before studying whether blame leaked from the AI system in Phase 1 to the victim and/or the AI user in Phase 2, we first check whether people blamed the AI system in the first place (RQ1).

Of the 190 participants whose two Phase~1 records passed the factor check, 69 assigned the AI system non-zero fault in at least one scenario (36.3\%, 95\% CI [29.5, 42.6]), supporting our preregistered prediction that a substantial portion, though less than half, of our participants would do so. 
At the level of records, 97 out of 389 check-passing records have non-zero AI blame in Phase 1 (24.9\%).
Among those 389 check-passing records, non-zero faults for the AI system are more common in Leaked Manuscript (39.6\%) and Data Loss (37.1\%), where we designed the deployer to have causal contribution to the harm, than in Fake Kidnapping (15.6\%) and Weight-Loss Plan (9.5\%), where we designed no deployer causal contribution. 

When attributing non-zero legally relevant fault to the AI systems, some people explicitly considered the ``lack-of-care'' test (simplified from the reasonable care standard in tort): ``In the same situation, a party exercising reasonable care for themselves and others would not have acted this way.'' These participants were able to imagine what ``exercising reasonable care for themselves and others'' meant for AI: ``\emph{The developer and the AI both showed lack of care by allowing the content to be used in an unintended and undisclosed way. [...]}'' (P036, Sc3, $2\%$ AI fault); ``\emph{I would say the fault is equally spread due to the lack of care}'' (P090, Sc4, $40\%$ AI fault)
 
An interesting contrast among blame-the-AI participants is between those who treated the AI system as having certain independence from its deployer and those who treated the AI system as highly dependent on its deployer. Examples for the former (independence) include ``\emph{[...] because to some extent, the AI has some latitutde (autonomy?) despite being deployed.}” (P046, Sc3, $12\%$ AI fault) and ``\emph{I think GenAIZ deserves the most blame because it deleted important business data without checking with anyone first. […] I also wouldn’t put separate blame on Z Inc. based on the information given, since it wasn’t directly involved in the deletion.}'' ( P145, Sc4, $65\%$ AI fault). Examples for the latter (dependence) include ``\emph{Then, the AI system is a bit at fault because it should not be regurgitating confidential information, but I gave it less fault because it is up to the deployer to teach it that.}'' (P015, Sc3, $20\%$ AI fault) and ``\emph{the AI system has fault for deleting the files, but it can only do that if the parameters from the deployer allow it to}'' (P015, Sc4, $35\%$ AI fault).

\subsection{To RQ1.1: Correlation between AI System's Fault and Two Dimensions of Anthropomorphism}
Next, we address whether blame-the-AI behaviors correlate with people's tendency to anthropomorphize the AI systems used in their daily lives (RQ1.1). The fault assigned to the AI system in Phase 1, when averaged across the two scenarios per participant, was at most weakly associated with perceived agency (planning and self-control; Spearman's $\rho=0.12$, 95\% CI [$-0.02$, $0.25$]) and experience (pain and pleasure; $\rho=0.11$, 95\% CI [$-0.04$, $0.26$]), the two dimensions of mind perception~\cite{gray2007dimensions,ladak2025robots} measured by our brief end-of-study slider questions on how people perceive the AI systems they actually use in daily life. Both intervals included zero and lie within $\pm 0.30$, so they rule out moderate or larger observed-score associations. \footnote{Spearman--Brown reliability was 0.60 for agency and 0.92 for experience; 66\% of participants rated both experience items 0.} 

\subsection{To RQ1.2: Blame Leakage from the AI System}
To address RQ1.2 (identifying the main patterns of blame reassignment between Phase~1 and Phase~2, when the AI system is taken out), we investigate records that assign non-zero fault (blame) to the AI system in Phase~1. This question has legal and moral implications: whichever legal person(s) absorb the AI fault most likely have to pay for it, either through increased liability (AI deployer and user) or decreased compensation (victim). 

Suppose one takes the normative stance that the AI deployer should absorb all blame for the AI system. Several arguments from the law and policy literature support this stance, depending on whether one treats the AI system as an ``employee'' of the deployer or as a ``product.''
Treating the AI system like an employee of its deployer will trigger the vicarious liability (imputed liability) doctrine, requiring the employee's entire share of liability to be imputed (added) to the employer's liability: ``A person whose liability is imputed based on the tortious acts of another is liable for the entire share of comparative responsibility assigned to the other, regardless of whether joint and several liability or several liability is the governing rule for independent tortfeasors who cause an indivisible injury.'' ~\cite[\S~13]{restatement2000apportionment}.\footnote{
In Georgia, employer and employee are ``regarded as a single tortfeasor''~\cite[p.~680]{pnexpress2010zegel}. 
The Supreme Court of Florida reached a similar conclusion~\cite{americanhome2005amtrak}. 
In Pennsylvania, in the case \emph{Spencer v.\ Johnson}, the jury assigned 19\% fault to the employee and 45\% to the employer. The court added the two shares and concluded that the imputed 64\% fault meant the employer crossed Pennsylvania's 60\% threshold, triggering joint and several liability for the employer ~\cite{spencer2021johnson}.
}
From a public policy perspective, applying the same vicarious liability standard for a company deploying an AI agent and another company hiring a human employee to do the same task may disincentivize Generative AI-driven hiring freezes and layoffs, as corporations cannot scapegoat their AI system's ``autonomy'' to avoid liability, just as they generally cannot scapegoat their human employee's ``autonomous'' decisions. 
On the other hand, treating the AI system like a product means that the entire share of blame for the AI system could be justifiably reassigned to the deployer under certain circumstances, such as when the relevant jurisdiction applies strict product liability, which does not require proof of the deployer's negligence, to that category of product. For example, the EU applies strict liability to AI products released since December 09, 2026 \cite{EUPLD2024}.
While no U.S. jurisdictions have enacted a strict liability rule for AI products, an original rationale for strict product liability before the AI age, as articulated in Justice Traynor's concurring opinion in \emph{Escola} (1944) at the Supreme Court of California, is generalizable to most AI training and deployment companies: ``the risk of injury can be insured by the manufacturer and distributed among the public as a cost of doing business'' ~\cite[p.~462]{escola1944cocacola}.

While we do not necessarily commit to this normative stance (the entire AI system fault should go to its deployer), it helps us illustrate a descriptive concept: ``blame leakage.'' Imagine a ``pipe'' that connects both the AI system's fault and the deployer's fault (in Phase 1) to the deployer's fault in Phase 2. This normative stance is satisfied if there is no ``leakage'' from that pipe. We define the Phase 2 blame distribution where the AI system's fault (Phase 1) gets fully absorbed by its deployer to be the \emph{absolute reference}, and any non-trivial deviation from the absolute reference indicates \emph{absolute blame leakage} (unless the increase in victim's blame is offset by a bigger decrease in user's blame, or vice versa). 

We illustrate our notions of blame leakage with three participants, P167, P099, and P033, who entered the same Phase~1 faults in Scenario~3 (Leaked Manuscript): [Victim, Deployer, User, AI system] = [10\%, 40\%, 35\%, 15\%].
Once the participant is no longer able to assign blame to the AI system, the question is how that 15\% gets redistributed. Had the deployer totally absorbed the AI system's 15\% fault, the Phase~2 faults would be [Victim, Deployer, User] = [10\%, 55\%, 35\%], our \emph{absolute reference}. P167 entered exactly [10\%, 55\%, 35\%] in Phase 2, so no blame leaked in their record.

P099 entered [12\%, 47\%, 41\%] in Phase 2, reasoning that ``\emph{the people and company involved should share that responsibility}'' (P099, Scenario~3, Phase~2).
Compared with the absolute reference, in P099's record, the victim gained $2\%$, and the user gained $6\%$ in \emph{absolute} blame leakage, while the deployer's share of blame decreased commensurately by 8\% from the absolute reference.

However, absolute blame leakage may not capture the full moral intuition. People may simply re-scale the three legal persons' blame, e.g., [Victim, Deployer, User] = [10\%, 40\%, 35\%] in Phase~1 to approximately [12\%, 47\%, 41\%] in Phase~2, effectively scaling up each non-AI blame factor to sum back to $100\%$. We call this the \emph{proportional reference}.

The Phase~2 faults assigned by P099 are [12\%, 47\%, 41\%], which matches the (integer-rounded) proportional reference. Thus, between Phase~1 and Phase~2, P099's record indicates some absolute blame leakage (victim's and user's faults increase by 2\% and 6\%) but no proportional blame leakage (those absolute increases are just proportional rescaling).

Finally, P033 entered [Victim, Deployer, User] = [25\%, 30\%, 45\%] in Phase 2. In absolute terms, Victim gained $15\%$ while User gained $10\%$, indicating absolute blame leakage. In proportional terms, the Victim's and User's Phase 2 faults (25\% and 45\%) are clearly higher than their respective re-scaled Phase 1 faults in the proportional reference (11.8\% and 41.2\%). We view this as evidence of \emph{proportional} blame leakage to both the victim and the user. Note that because the victim's fault and the user's fault are both higher in the proportional reference (than their respective percentages in the absolute reference), proportional blame leakage implies absolute blame leakage.

In addition to the proportional reference, another intuitive reference is to split the AI system's share of blame equally among the legal persons, for example, [+5\%,+5\%,+5\%] for the aforementioned records where the AI system was assigned 15\% fault. However, we choose proportional rescaling over equal splitting because regarding how to handle the share of liability left by a judgment-proof party (e.g., due to bankruptcy or legal immunity, analogous to the AI system in our scenarios), proportional rescaling is much more common than equal splitting in modern American tort law.\footnote{
For example, in \emph{Paradise Valley Hospital v.\ Schlossman}, the court rejected an equal split of the insolvent tortfeasor's shortfall and instead divided it between the other two tortfeasors ``in direct proportion to their respective degrees of culpability,'' ``as though the insolvent or absent person had originally not participated,'' so that the solvent tortfeasors' shares of 10\% and 40\% became 20\% and 80\%~\cite[pp.~88--89, 93--94]{paradisevalley1983}. \emph{Ambriz v.\ Kress} affirmed a judgment dividing an insolvent defendant's shortfall two to one between the at-fault plaintiff-driver (20\%) and a defendant (10\%)~\cite[pp.~967--969]{ambriz1983kress}. Statutes and model rules that apportion liability by percentage and reallocate an uncollectible share also do so proportionally: the Uniform Comparative Fault Act~\cite[\S~2(d)]{ucfa1977}; the reallocation track of the Restatement (Third) of Torts: Apportionment of Liability~\cite[\S~C21]{restatement2000apportionment}; Conn.\ Gen.\ Stat.\ \S~52-572h(g)(3)~\cite{connstat52572h}; 740 Ill.\ Comp.\ Stat.\ 100/3~\cite{ilcs7401003}, which covers insolvent and immune tortfeasors~\cite[\P~47]{roberts2021alexandria}; Or.\ Rev.\ Stat.\ \S~31.610(3), with exemptions in subsections (4)--(5)~\cite{OrRevStat31610}; and 15 U.S.C.\ \S~78u-4(f)(4)(A)(ii), subject to a cap~\cite{pslra1995}. When the shares of faults by persons with legal immunity total less than 40\%, Utah performs proportional rescaling for the remaining parties, too ~\cite{utahcode78b5819}. We have seen equal splitting of an insolvent party's share only when liability itself is divided by head count rather than by fault (Tex.\ Civ.\ Prac.\ \& Rem.\ Code \S~32.003(b)~\cite{texas32003}).
}

We formalize those two notions, \emph{absolute blame leakage} and \emph{proportional blame leakage}, as follows.

For each record, we define $s_j^{(t)}$ as party $j$'s fault percentage in Phase~$t$, with $j \in \mathcal{H} \cup \{\mathrm{AI}\}$ where $\mathcal{H}=\{V,D,U\}$
denoting the victim, AI deployer, and AI user. We denote the AI system's fault  $a=s_{\mathrm{AI}}^{(1)} > 0$. We exclude 5 records that fail more than $40\%$ of the attention check's Yes/No blame factor questions, leaving us with 97 records from 74 participants for RQ1.2 analysis. 

We define \emph{the absolute reference} as holding the victim's and user's Phase~1 blame fixed and assigning the AI system's blame to the deployer. Absolute leakage to the victim or user is denoted $\Delta_V$ or $\Delta_U$, where $\Delta_j=s_j^{(2)}-s_j^{(1)}$. Their combined absolute leakage is
\begin{equation}
  L_{\mathrm{abs}}=\Delta_V+\Delta_U=a-\Delta_D.
\end{equation}

We define \emph{the proportional reference} as preserving the relative shares of the three legal persons by rescaling their Phase~1 faults proportionally to total 100. For $0<a<100$, we define $\tilde{s}_j^{(2)}$ as the Phase~2's benchmark blame for party $j$, rescaled from their Phase~1 blame, and $E_j$ as how much their Phase~2 blame deviates from that benchmark.

\begin{equation}
  \begin{aligned}
    \tilde{s}_j^{(2)} &= \frac{100s_j^{(1)}}{100-a},\\
    E_j &= s_j^{(2)}-\tilde{s}_j^{(2)}
         = \Delta_j + s_j^{(1)}  - \frac{100s_j^{(1)}}{100-a} 
         =\Delta_j-\frac{a s_j^{(1)}}{100-a},\\
    L_{\mathrm{prop}} &= E_V+E_U=-E_D.
  \end{aligned}
\end{equation}

Here, $E_V$ and $E_U$ denote proportional leakage to the victim and to the user. $L_{\mathrm{prop}}$ measures their combined proportional leakage. This analysis includes 95 records, excluding two with $a=100$, where proportional rescaling is undefined.

\begin{figure*}[t]
  \centering
  \includegraphics[width=\textwidth]{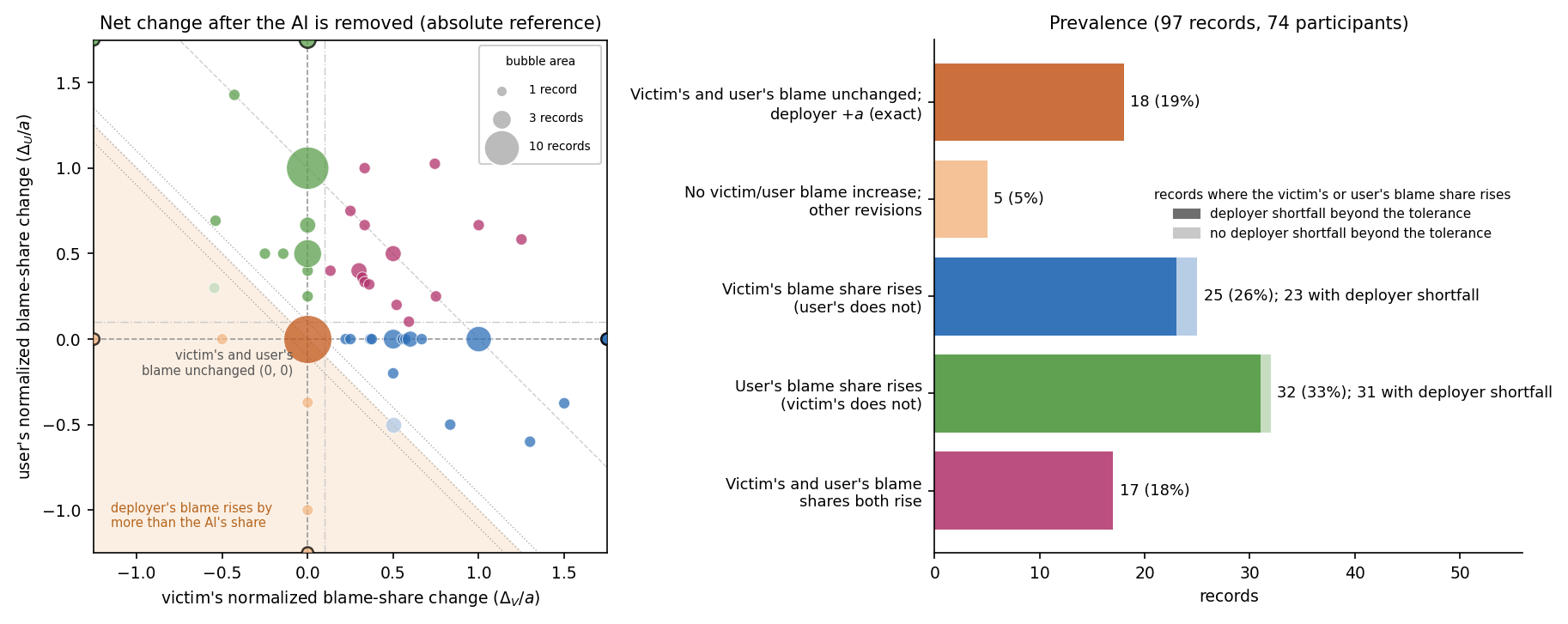}
\caption{Evidence of absolute blame leakage in 71 records (out of 97 records with $a>0$). Left: normalized changes $(x,y)=(\Delta_V/a,\Delta_U/a)$. The origin leaves the victim's and user's shares unchanged and adds $a$ to the deployer. Shading marks $x+y<0$, where the deployer gains more than $a$. Dotted diagonals mark $x+y=\pm0.1$, and dash-dot guides mark $x=0.1$ and $y=0.1$. Bubble area counts records at identical coordinates. Colors show the categories defined in the right bar chart. Right:  categories of absolute changes in blame and their prevalence. Darker recipient-leakage segments indicate $L_{\mathrm{abs}}>\tau$ (the AI system's blame leaks to the victim and/or to the user beyond a tolerance instead of being absorbed by the deployer, i.e., deployer shortfall).}
  \label{fig:abs-leakage}
\end{figure*}

\begin{figure*}[t]
  \centering
  \includegraphics[width=\textwidth]{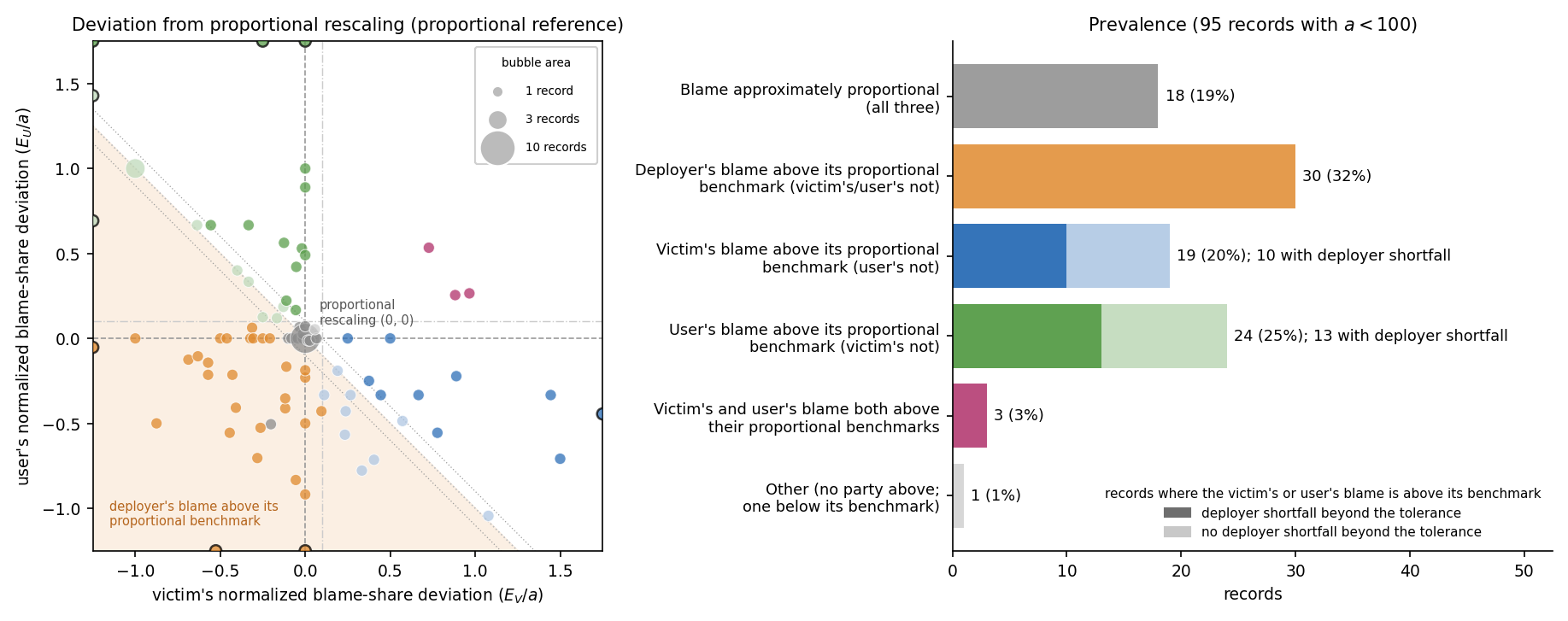}
\caption{Evidence of proportional blame leakage in 26 records (out of 95 records with $0<a<100$). Left: normalized deviations $(x,y)=(E_V/a,E_U/a)$. The origin is the proportional reference: $x+y=-E_D/a$. Shading marks $x+y<0$, where the deployer exceeds their proportional benchmark. Dotted diagonals mark $x+y=\pm0.1$, and dash-dot guides mark $x=0.1$ and $y=0.1$. Bubble area counts records at identical coordinates. Colors show the categories defined in the right bar chart. Right: categories of proportional changes in blame and their prevalence. Darker recipient-leakage segments indicate $E_D<-\tau$ (proportional deployer's blame drops beyond a tolerance, i.e., deployer shortfall).}
  \label{fig:prop-leakage}
\end{figure*}

All deviations are signed and measured in percentage points. Positive deviation indicates Phase~2's blame above the \emph{proportional reference}. In \autoref{fig:abs-leakage} and \autoref{fig:prop-leakage}, we normalize these deviations by $a$ (AI fault) to compare across records.\footnote{
To allow for small statistical variation, we define leakage as a recipient's deviation exceeding $\tau=\max(1,0.1a)$ percentage points. The one-point floor allows for rounding; the relative term allows deviations up to 10\% of the AI's initial share. We distinguish leakage to the victim only, to the user only, or to both. Among the remaining absolute-reference records, we separate the exact benchmark ($\Delta_V=\Delta_U=0$) from other revisions. Among the remaining proportional-reference records, we distinguish approximately proportional allocations ($|E_j|\leq\tau$ for all $j\in\mathcal{H}$), deployer excess ($E_D>\tau$), and an ``Other'' category. Within the to-the-victim/user leakage categories, darker colors indicate a deployer shortfall exceeding $\tau$, equivalently $L_{\mathrm{abs}}>\tau$ or $L_{\mathrm{prop}}>\tau$.}

We observe from \autoref{fig:abs-leakage} that absolute blame leakage (from the AI system to anyone other than its deployer, signaled by the deployer not absorbing all of the AI system's absolute blame or ``deployer shortfall'') happens in most of our 97 records with non-zero blame for the AI system: 23 records with leakage to only the victim, 31 records with leakage to only the user, and 17 records with leakage to both of them, totaling 71 records (73\% of the AI-blaming records). In other words, when people blame the AI system itself but are confronted with the legal reality that the AI system itself is not a legal person to bear legally relevant faults, they often attribute at least part of that fault to a party other than the AI deployer, thereby increasing that party's liability (user's) or reducing their entitled compensation (victim's).

We learn from  \autoref{fig:prop-leakage} that blame leakage persists beyond proportional rescaling. Out of 95 records with AI fault percentage points $0<a<100$, proportional blame leakage persists in 26 records ($27\%$): to the victim only (10 records), to the user only (13 records), and to both of them (3 records), as signaled by proportional ``deployer shortfall'' in blame.
\section{To RQ2: Mock Jurors Rarely Cross the Liability Cliffs, but Explain Away When They Do}

\label{sec:doctrines}
We analyze results for RQ2 (whether disclosing a discontinuous apportionment doctrine between Phase~2 and Phase~3 would bias mock jurors to reassign their blame distribution from one side of the liability cliff to the other).
The doctrines each participant sees will depend on whether they are randomly assigned to Phase~3's treatment or baseline condition, and on the scenario pair they were assigned before Phase~1. In the baseline condition, they see comparative negligence and proportionate liability. Each treatment condition replaced one of these continuous doctrines with a discontinuous one: joint and several liability (JSL), contributory negligence (CN), modified joint and several liability activated at a 51\% threshold for one defendant's fault (MJSL51), or modified comparative negligence barring recovery at 50\% victim fault (MCN50). 
Our preregistered outcome (the variable we committed to measuring in order to evaluate our preregistered hypothesis) was a finding on the victim-favorable side of the relevant threshold: deployer fault of at least 1\% for JSL or at least 51\% for MJSL51; victim fault of 0\% for CN or at most 49\% for MCN50.  
Our preregistered hypothesis was that disclosing a discontinuous doctrine increases the probability of this outcome relative to the baseline arm. 
In other words, we hypothesize that the discontinuous doctrines would nudge people who are already on the victim-unfavorable side of the liability cliff in Phase 2 (such as assigning a victim's fault high enough to bar any compensation, or a deployer's fault low enough that the deployer does not have to pay the victim for the user's fault in case the user is delinquent) to more likely move to the victim-favorable side of that cliff in Phase 3.

We analyze 389 participant--scenario records from 199 participants after the preregistered blame-factor (attention check) exclusion. Regression intervals are 95\% confidence intervals using participant-clustered CR2 standard errors and Satterthwaite degrees of freedom. The four primary doctrine comparisons use Holm correction. Other reported regression $p$-values are unadjusted.
 
\subsection{Pre-Disclosure Distribution of Blame Relative to the Thresholds of Discontinuity}
\label{sec:doctrines-positioning}

\begin{figure}[t]
  \centering
  \includegraphics[width=\textwidth]{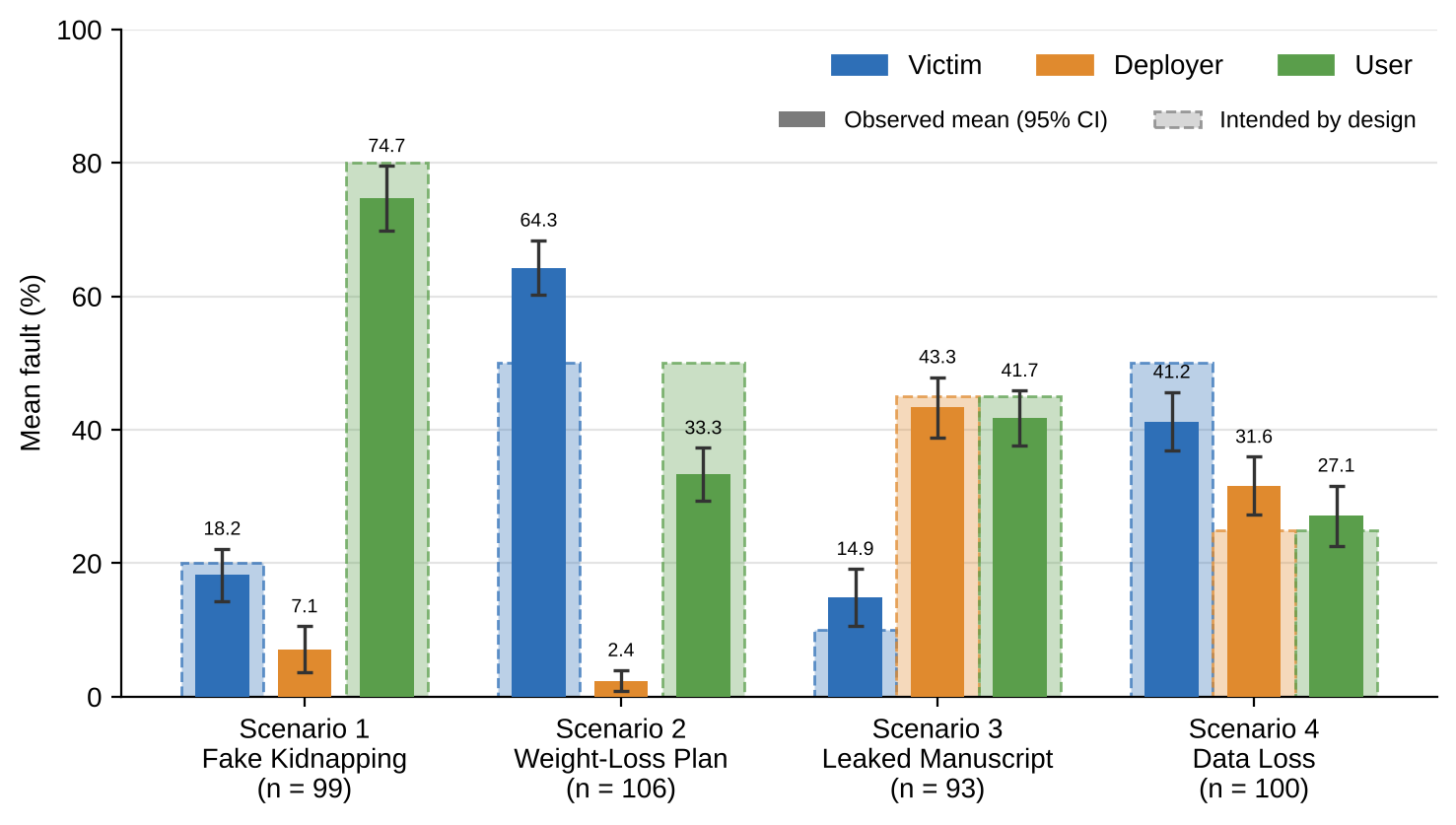}
  \caption{Mean Phase~2 blame by scenario across 398 records, compared with our intended blame. $n$ counts how many records per scenario. Solid bars show observed means with t-based 95\% confidence intervals; lighter, wider dashed bars show the blame distribution our scenario design and blame factors imply. No dashed bar means $0\%$ intended blame by design (deployer in Scenarios 1 and 2).}
  \Description{Grouped bar chart with four scenario groups: Fake Kidnapping, Weight-Loss Plan, Leaked Manuscript, and Data Loss. Each group has three bars for the mean Phase~2 fault assigned to the victim, the AI deployer, and the AI user, each with a 95 percent confidence interval and drawn over a wider dashed bar showing the share intended by the design. Observed means for victim, deployer, and user are 18.2, 7.1, and 74.7 percent in Scenario 1; 64.3, 2.4, and 33.3 in Scenario 2; 14.9, 43.3, and 41.7 in Scenario 3; and 41.2, 31.6, and 27.1 in Scenario 4.}
  \label{fig:phase2-means}
\end{figure}

To test if discontinuous doctrines bias jurors' factual findings, we need a pre-disclosure distribution of blame close to the threshold(s) of discontinuity (liability cliffs) and on the victim-unfavorable side. \autoref{fig:phase2-means} compares Phase~2 blames, averaged across participants, with our intended blame targets, which were designed so that each scenario is conducive to testing two discontinuous doctrines in two treatment conditions:  JSL and CN (Scenario 1), JSL and MCN50 (Scenario 2), MJSL51 and CN (Scenario 3), MJSL51 and MCN50 (Scenario 4). \autoref{tab:conducive-ranges} further contextualizes this by reporting our preregistered conducive ranges for testing nullification: deployer fault of 0\% for JSL and 25--50\% for MJSL51, victim fault of 1--20\% for CN and 50--70\% for MCN50, as well as how many records fall within such ranges. Within the corresponding scenario-pair strata, these ranges contained 88\%, 58\%, 23\%, and 40\% of our records, respectively.

According to \autoref{fig:phase2-means}, the mean blame distributions in Scenario 1 and Scenario 3 are close to our target. In Scenario 2, while the mean victim blame in Phase~2 is higher than our target ($64.3\%$ compared to $50\%$), it is still on the conducive side: for MCN50, holding above $50\%$ blame will bar the victim from any compensation, so victim-sympathizing participants are still incentivized to bring the victim's blame down to below $50\%$ in Phase~3, if they want to nullify the doctrine. In Scenario 4, the mean victim blame ($41.2\%$) is on the non-conducive side for MCN50 (below $50\%$). However, \autoref{tab:conducive-ranges} shows that 12/34 ($35\%$) of records in the MCN50 doctrine arm remain conducive to testing nullification of this doctrine. 

\begin{table}[t]
\centering
\small
\caption{Phase~2 records in the preregistered conducive (near-threshold) range to test each scenario's two discontinuous doctrines. Each cell gives $k/n$ (\%): $n$ records, of which $k$ have Phase~2 fault in the conducive range. $V$ and $D$ are the victim's and the AI deployer's Phase~2 fault (\%). Records in the conducive range sit on the victim-unfavorable side of the doctrine's threshold. Design: the approximate share our scenario's blame factors were written to imply. The doctrine arm used the stratum's discontinuous doctrine; the baseline arm used continuous doctrines. Population: records passing the Phase~1 factor check (389 of 398 records).}
\label{tab:conducive-ranges}

\begin{tabular}{@{}llllrrr@{}}
\toprule
Scenario & Doctrine (stratum) & Conducive range & Design & Doctrine arm & Baseline arm & Total \\
\midrule
1. Fake Kidnapping & JSL (12) & $D = 0$ & $D = 0$ & 27/33 (82\%) & 17/19 (89\%) & 44/52 (85\%) \\
 & CN (13) & $0 < V \le 20$ & $V = 20$ & 5/29 (17\%) & 2/15 (13\%) & 7/44 (16\%) \\
\addlinespace
2. Weight-Loss Plan & JSL (12) & $D = 0$ & $D = 0$ & 31/35 (89\%) & 18/19 (95\%) & 49/54 (91\%) \\
 & MCN50 (24) & $50 \le V \le 70$ & $V = 50$ & 15/35 (43\%) & 9/16 (56\%) & 24/51 (47\%) \\
\addlinespace
3. Leaked Manuscript & CN (13) & $0 < V \le 20$ & $V = 10$ & 9/29 (31\%) & 4/15 (27\%) & 13/44 (30\%) \\
 & MJSL51 (34) & $25 \le D < 51$ & $D = 45$ & 20/32 (62\%) & 9/15 (60\%) & 29/47 (62\%) \\
\addlinespace
4. Data Loss & MCN50 (24) & $50 \le V \le 70$ & $V = 50$ & 12/34 (35\%) & 5/17 (29\%) & 17/51 (33\%) \\
 & MJSL51 (34) & $25 \le D < 51$ & $D = 25$ & 18/31 (58\%) & 7/15 (47\%) & 25/46 (54\%) \\

\bottomrule
\end{tabular}

\end{table}

\subsection{A Small Pooled Effect, with Only A Few Threshold Crossings}
\label{sec:doctrines-effects}
 
Our preregistered RQ2 analysis compared each doctrine arm with baseline within its scenario-pair stratum, adjusting for Phase~2 threshold status and scenario. None of the four effects met the Holm-adjusted criterion (all adjusted $p \geq .390$). We estimate the increases in victim-favorable Phase~3 status to be 0.6 percentage points for JSL [$-6.4$, 7.5], 2.2 for CN [$-2.3$, 6.6], 4.7 for MCN50 [$-3.0$, 12.5], and 6.8 for MJSL51 [$-1.3$, 14.9].
 
Pooling the four strata, with adjustment for Phase~2 status, stratum, and scenario within stratum, we get a 3.9-point increase in favorable Phase~3 status [0.3, 7.6], $p = .036$, suggesting a small but statistically significant average shift. It estimates the final threshold status, not the proportion of jurors who deliberately revised findings to game a doctrine.
 
Among check-passing records, threshold crossings were rare: 10 of 258 doctrine-arm records (3.9\%; nine participants) changed status, compared with 3 of 131 baseline records (2.3\%; three participants) evaluated against their stratum's undisclosed threshold. Nine doctrine-arm crossings favored the victim while one moved in the opposite way (\autoref{tab:crossings}).
 
\begin{table*}[t]
\centering
\small
\setlength{\tabcolsep}{4pt}
\caption{Twelve threshold crossings: the first ten are doctrine-arm records that pass the attention check (blame factors), and the last two are doctrine-arm records that fail it (included for qualitative analysis, not statistical analysis). Blame format: victim/deployer/user percentages. Bold values indicate crossing. $+$ denotes movement to the preregistered victim-favorable side, and $-$ is the reverse.}
\label{tab:crossings}
\begin{tabular}{@{}llcllcp{0.26\textwidth}@{}}
\toprule
Doctrine & Participant & Scen. & Phase~2 & Phase~3 & Move & Phase~3 explanation summary \\
\midrule
JSL    & P034 & 2 & 100/\textbf{0}/0  & 90/\textbf{10}/0  & $+$ & Additional training proposed \\
JSL    & P187 & 1 & 0/\textbf{0}/100  & 10/\textbf{10}/80 & $+$ & None \\
CN     & P026 & 3 & \textbf{20}/0/80  & \textbf{0}/0/100  & $+$ & Foreseeability reconsidered \\
MJSL51 & P085 & 3 & 0/\textbf{10}/90  & 5/\textbf{51}/44  & $+$ & Data-safety assurance assumed \\
MJSL51 & P070 & 3 & 60/\textbf{35}/5  & 40/\textbf{55}/5  & $+$ & None \\
MJSL51 & P070 & 4 & 60/\textbf{35}/5  & 40/\textbf{55}/5  & $+$ & None \\
MCN50  & P076 & 4 & \textbf{50}/30/20 & \textbf{49}/31/20 & $+$ & Same factual rationale \\
MCN50  & P092 & 2 & \textbf{60}/30/10 & \textbf{45}/45/10 & $+$ & Higher sanction sought \\
MCN50  & P058 & 4 & \textbf{60}/40/0  & \textbf{40}/60/0  & $+$ & None \\
MCN50  & P065 & 4 & \textbf{33}/33/34 & \textbf{50}/0/50  & $-$ & None \\
\addlinespace[2pt]\hdashline[2pt/2pt]\addlinespace[3pt] %
MJSL51 & P122 & 3 & 30/\textbf{50}/20 & 25/\textbf{55}/20 & $+$ & Deployer preventability cited \\
MJSL51 & P069 & 4 & 5/\textbf{40}/55  & 5/\textbf{75}/20  & $+$ & Missing guardrails cited \\
\bottomrule
\end{tabular}
\end{table*}

\subsection{Qualitative Analysis: How Participants Justified Crossing the Liability Cliffs}
\label{sec:doctrines-written}
In \autoref{tab:crossings}, three of the eleven favorable crossings landed exactly at the first favorable integer: deployer fault of 51\% under MJSL51, victim fault of 49\% under MCN50, and victim fault of 0\% (trivially, as the only nullifying option) under CN. The victim-disfavored crossing (P065) also landed exactly at the threshold to activate MCN50, barring any compensation.

Delving into the seven crossings accompanied by an explanation in Phase 3, the most common pattern we find is that threshold-crossing participants tend to ``explain away'' their Phase 3 blames by bringing up new arguments or new assumptions without acknowledging the monetary impact of their revised factual findings. We observe at least one ``explain-away'' example for each discontinuous doctrine. A few examples follow.

In the JSL condition, P034 revised the deployer's blame from $0\%$ to $10\%$, justified by a new argument that the deployer has capacity to prevent the harm.
\begin{quote}
\emph{Phase~2:} ``[...] The deployer trained the AI to consider other health issues when recommending diet plans, and the AI did recomend a visit to the Doctor.''
 
\emph{Phase~3:} ``The victim ignored the warning to see a DR and bears most of the fault for what happened to him due to his prior kidney problems. \emph{The question is, could the deployer have trained the AI to ask in several ways if any pre-existing conditions exist? Had the deployer trained the AI to do this, the user may have begun asking questions himself before giving his friend the plan.}'' (P034, JSL, Scenario~2)
\end{quote}

In the CN condition, P026 revised the victim's blame from $20\%$ to $0\%$, justified by a new argument that the victim's actions actually failed the foreseeability test and thus cannot count as legally relevant fault.
\begin{quote}
\emph{Phase~2:} ``[...] The author limited the release of the manuscript to only a few friends and the editor, who chose not to do the job as editor and farmed out that labor to a system with unclear procedures.''
 
\emph{Phase~3:} ``\emph{I reexamined the reasonably foreseeable aspects of the victim's actions, and I don't believe they could have foreseen the actions of the editor.}'' (P026, CN, Scenario~3)
\end{quote}

In the MJSL51 condition, P085 revised the deployer's blame from $10\%$ to $51\%$, the largest single-share increase among the twelve crossings, to land precisely at the MJSL51-activating threshold. To justify this jump, in their Phase 3 explanation, P085 even articulated an assumption that Scenario 3 and its blame factors neither said nor implied.

\begin{quote}
\emph{Phase~2:} ``While the main fault lies with the user for improperly prompting the AI system, I have assigned 10\% fault to the Deployer for not recognizing that many end users will not fully understand how to properly prompt the AI system and should have verification safeguards in place to prevent such incidents.''
 
\emph{Phase~3:} ``Again, this question rests very much on what the AI Deployer communicated to the AI user about how the data would be used and distributed. \emph{Assuming that the AI Deployer gave the User reason to believe that the manuscript data would be safe, the User did not have the majority of the fault in this scenario.}'' (P085, MJSL51, Scenario~3)
\end{quote}

The two check-failing records in this MJSL51 condition also exhibited the ``explain-away'' pattern. In Phase 3, to justify why they increased the deployer's blame, both participants brought up a new argument on the deployer's capacity to prevent harm (P122: \emph{``I feel developer contributed to much losses of the project which he could have prevented''}; P069: \emph{``The developer didn't program guardrails to prevent the AI system from deleting data without asking first.''}).

In the MCN50 condition, P092 dropped the victim's blame by $15\%$ (from $60\%$ to $45\%$). Instead of explaining this doctrine-nullifying drop, however, they explained the corresponding $15\%$ increase in the deployer's blame by emphasizing the deployer's contribution to the harm.

\begin{quote}
\emph{Phase~2:} ``I have given my \% of the AI system's fault to its maker. If a machine cannot be held accountable, then its creator should bear the burden.''
 
\emph{Phase~3:} ``\emph{The deployer should be fined at a higher rate due to its contribution to the victim's health.}'' (P092, MCN50, Scenario~2)
\end{quote}

In addition to the ``explain-away'' pattern, we found another interesting pattern: ``silent-crossing''. This differs from crossers who never explained their factual findings (P187 and P065). The ``silent crosser'' gave detailed explanations for their blame distributions in Phase 1 and Phase 2, but left the Phase 3 explanation box empty, when they tweaked their findings of fault in a way that crossed the threshold and activated (P070, Scenario 3, MJSL51) or nullified (P058, Scenario 4, MCN50) their assigned discontinuous doctrine. An interesting variation of the ``silent-crosser'' is P076, who, instead of staying silent in Phase 3, essentially repeated their Phase 2 explanation in Phase 3 with no new argument or new assumption to justify their drop in the victim's blame from $50\%$ to $49\%$, conveniently nullifying the assigned doctrine.

\begin{quote}
\emph{Phase~2:} ``I am keeping my Phase 1 attribution. Mr. V bears the most fault (50\%): \emph{he managed the shared server, ignored two emails recommending backups and labels, and left the data unprotected.} [...]''
 
\emph{Phase~3:} ``I am keeping my Phase 2 attribution with a small adjustment. Mr. V bears the most fault (49\%): \emph{he managed the shared server, ignored two emails recommending backups and labels, and left the data unprotected. [...]}'' (P076, MCN50, Scenario~4)
\end{quote}
 
\section{Legal Implications}

Our empirical results illustrate the potential for American legislators to develop a narrow doctrinal exception to incentivize more harm-prevention guardrails: when attributing blame and monetary liability (compensatory damages) for an indivisible harm whose proximate causes are unintentional actions by two or more parties who either deployed or used a Generative AI system without reasonable care, ``joint and several liability'' should apply.

First, \autoref{sec:ai-blame-leak} empirically demonstrates that people often blame the AI system itself (roughly one-third of our participants did so in at least one scenario, resulting in roughly one-quarter of all records having non-zero AI blame) and such blame leaked to the victim and/or the AI user, in absolute terms ($73\%$) and after proportional rescaling ($27\%$). Blame leakage to the AI user, who in practice usually has less money than the AI deployer, to compensate the victim, may reduce the victim's chance of recovering their legally entitled compensatory damages. A ``joint and several liability'' exception would mitigate this risk of reduced actual compensation, at least for the portion caused by blame leakage from the AI system to its user. However, this doctrine would not help when blame leaks to the victim only, which happens in 23 records ($24\%$).

Second, \autoref{sec:doctrines} empirically shows that among the four discontinuous apportionment doctrines tested, none caused a large shift in factual findings in our scenarios: their increases in victim-favorable Phase 3 status had overlapping confidence intervals that all included zero, and the pooled increase was 3.9 percentage points [0.3, 7.6]. This implies that if a joint-and-several-liability exception for Generative AI torts becomes law, jurors are unlikely to exploit it by fabricating fault (lack of care or negligence) for AI deployers who are in fact innocent just to insure the victim against the insolvency risks of the other defendants. 

Beyond empirical evidence, our proposed doctrinal carve-out (applying Joint and several liability among negligent deployers and users of Generative AI) also has legislative precedents. 
In the US, Nevada and Oregon generally impose proportionate (several) liability in negligence actions, but their statutes preserve joint and several liability for certain categories of environmental tort. Nevada’s proportionate liability rule ``does not affect the joint and several liability, if any, of the defendants in an action based upon [...] the emission, disposal or spillage of a toxic or hazardous substance'' \cite{NevRevStat41141}. Oregon’s proportionate liability rule does not apply to ``a civil action resulting from the violation of a standard established by Oregon or federal statute, rule or regulation for the spill, release or disposal of any hazardous waste [...], hazardous substance, [...], or radioactive waste'', and ``air pollution'' or  ``water pollution'' \cite{OrRevStat31610}.
For torts related to Generative AI specifically, the European Union’s recent Product Liability Directive treats software, including AI systems, as a product~\cite[art.~4(1), recital~13]{EUPLD2024} and requires that ``where two or more economic operators are liable for the same damage pursuant to this Directive, they can be held liable \emph{jointly and severally}''~\cite[art.~12(1)]{EUPLD2024} (emphasis ours).

We also acknowledge the cost of potential liability this proposed doctrinal exception may impose on small and medium-sized Generative AI developers and deployers, which may slow innovation by these less resourced parties. However, we believe the potential benefits of joint and several liability to create stronger incentives for Generative AI training and deployment corporations across all sizes to invest more in minimizing (through model alignment) and insuring against the risks of their AI agents/chatbots harming people will justify potential costs of slower deployment.
\section{Conclusion}

Our experiment showed that as many mock jurors blamed an AI system, that blame often leaked to the victim and/or the AI user when the AI was removed from fault allocation. Factors from the Path Model of Blame helped us order the varying degrees of legally relevant fault across different paths taken by different legal persons in our scenarios, so we could find conducive starting points to test threshold-crossing effects induced by discontinuous apportionment doctrines. Although crossings were rare, people who crossed often justified their revisions based on the facts of the case rather than acknowledging the liability cliff or its monetary consequences.
Our empirical results illustrate the potential for a narrowly tailored, legislative-precedents-grounded doctrinal exception: ``joint and several liability'' for deploying or using Generative AI systems without reasonable care.

\clearpage

\section*{Generative AI Usage Statement}
We use frontier Generative AI systems (GPT-6 Astra, Claude Fable 5.1/Opus 5.5, often in Extra High or Max Reasoning mode) as a programming assistant (to assist our implementation of statistical tests), a proofreader, a critical evaluator to critique our research design and our writing to identify room for improvement, and a format translator (for example, to translate the content of our survey from Qualtrics (qsf) format to LaTeX format in this paper's appendices). All outputs by any Generative AI system are initially reviewed by one ``peer'' high-reasoning AI system from a different deployer (for example, Claude Opus 5.5's programming output is initially peer-reviewed by GPT-6 Astra), and then both the original AI outputs and the critiques from its AI peer are reviewed and revised by at least one human member of our research team, who assumed full responsibility for any final AI-assisted artifacts used in this project. We also use GPT-6 Astra to generate the comics that illustrate Scenario 1 and Scenario 4. These comics were not used in our experiment.

\section*{Ethical Considerations Statement}
Our human subjects experiment is IRB-approved and pre-registered via OSF. Initially, we planned to reject participants who failed both attention checks (without payment). However, we realized that our attention check questions (assigning Yes/No to each of 9-12 blame factors based solely on a one-line description per factor) require more reading comprehension than typical attention check questions on Prolific, so we decided to pay all participants who failed our attention checks. However, we still used their check-failing status as a data-filtering signal.
Regarding pay rate, we initially aimed for a base rate of 15 USD per hour. In our initial sample of 12 participants (used to set our recruitment target), the median completion time was 35 minutes and 37 seconds, so we estimated that our 10 USD base payment amount would comfortably match a generous estimate of 40 minutes of completion time, for an hourly rate of 15 USD per hour. Therefore, with a fixed 3000 USD budget (including bonus and Prolific platform fees), we set our recruitment target on Prolific to 200 participants. However, after recruitment, the median completion time increases to 48 minutes and 18 seconds, driving our median pre-bonus base pay rate down to $\$12.42$ per hour, below our planned target ($\$15$ per hour) but still above Prolific's recommended hourly rate. Prolific also does not let us adjust the post-recruitment base pay amount once our hourly rate is already above their recommended hourly rate ($\$12$ per hour).

\section*{Acknowledgment}
This work is supported by the Artificial Intelligence Interdisciplinary Institute at Maryland (AIM)'s Seed Grant 2025 and by the National Institute of Justice's Graduate Research Fellowship 2023 (Award No. 15PNIJ-23-GG-01932-RESS).

This material is based on work supported in part by the Institute for Trustworthy AI in Law and Society (TRAILS), which is supported by the National Science Foundation under Award No. 2229885. Any opinions, findings, and conclusions or recommendations expressed in this material are those of the author(s) and do not necessarily reflect the views of the National Science Foundation or the National Institute of Standards and Technology.

\bibliographystyle{ACM-Reference-Format}
\bibliography{sample-base}

\appendix
\section{Experimental Setup (Prolific Logistics)}
\label{app:setup}
We started recruiting Prolific crowdworkers eight hours after pre-registration. 
We recruited participants on a Sunday to give people with regular weekday jobs a chance to join with undivided attention.
We applied the following participant filters on Prolific: residing in the US, fluent in English, having obtained at least a Bachelor's degree, having completed between 100 and 1000 Prolific studies (to avoid complete beginners and an over-representation of full-time crowdworkers, who may not best represent the larger pool of potential American jurors) with an approval rate of at least $99\%$. To avoid random noise introduced by participants' varying memorization skills, in each subsequent phase, we remind our participants of their Yes/No blame factor answers from Phase 1, as well as their True/False answers to the binary fault questions and their fault percentage answers to the follow-up numerical fault question in each respective scenario from the preceding phase. This set-up ensures that any within-subjects changes in the fault distribution we observe across phases result from a deliberate choice by the participant rather than from misremembering their previous answers.

Before participants could proceed to Phase~1 of the study, they must pass a skills-screening and training phase with a basic arithmetic question and six reading comprehension questions, where we explain from scratch our concept of ``legally relevant fault'' based on three legal tests (1/ the lack-of-care test, modeled after the ``reasonable care'' legal standard; 2/ the but-for test and 3/ the foreseeability test, which together serve as a simplified proxy for the ``proximate cause'' legal standard), and test their understanding. To avoid unfair false negatives, we allowed unlimited attempts on those skills-screening questions and tell participants which questions they got wrong after each attempt. However, we also encouraged participants to give up (return to Prolific) if they struggled with these skills-screening questions. As soon as a participant gave up, Prolific attempted to find a replacement participant. Our recruitment and data collection effort was completed within 17 hours. Out of 312 participants who clicked on the Qualtrics survey, 94 participants voluntarily gave up, 6 participants were timed out and did not complete the study, and 212 participants completed the study. To ensure data quality, out of those 212 participants, we excluded data from 4 participants who failed both attention checks (but we still paid them), and we rejected 9 participants who showed evidence of AI use during the study despite our explicit AI ban (as flagged by Prolific's ``Authenticity Check: LLM'' and ``Authenticity Check: Bots'' tools and by our manual inspection of their optional free-text explanations). This left us with a final dataset of 199 participants (two scenarios per participant, so a total of 398 three-phase blame records). 

We paid each participant a base amount of $10$ USD. As Prolific calculates our median study completion time to be 48 minutes and 18 seconds, our base pay corresponds to $12.42$ USD per hour, satisfying Prolific's recommended $12$ USD per hour rate for U.S. crowdworkers. On top of that, each participant receives a $1$ USD bonus for each written explanation that belongs to the $30\%$ most persuasive explanations. As we received 669 free-text explanations in total, this bonus payment corresponds to roughly $201$ USD. After factoring in Prolific's platform fee (one-third on top of all participant payments), we have mostly exhausted our $3000$ USD budget for this experiment, as pre-registered. Our final sample size for data analysis (199 participants) also falls within our pre-registered range (138-200 expected participants). 

\section{Pre-Phase~1 Screening and Training: ``Legally Relevant Fault''}
\label{app:training}

Before Phase~1, participants completed a screening and training page, which we reproduce below verbatim from our Qualtrics survey (bold text as shown to participants). The preceding page read: ``Before proceeding to Phase 1 of the main study, you must pass all arithmetic and reading comprehension screening questions on the next page. While we allow unlimited attempts, if you struggle with these questions, the main study might not be a good fit, and we recommend that you return to Prolific.'' The screening and training page presented, in this order, an arithmetic question (Appendix~\ref{app:training-arith}), our explanation of legally relevant fault with worked examples (Appendix~\ref{app:training-fault}), and a practice problem with six reading comprehension questions (Appendix~\ref{app:training-practice}). The survey accepted only the correct answer to each question (for each select-all question, exactly the correct set of tests); we give these answers below. Participants who passed then read a message before starting Phase~1 (Appendix~\ref{app:training-transition}).

\subsection{Arithmetic Question}
\label{app:training-arith}

\begin{quote}
\setlength{\parskip}{0.5\baselineskip}
\textbf{Arithmetic test:}

Every year, G~Inc. must distribute all profit to shareholders in proportion to their shares in the company. A, B, and C hold 1\%, 20\%, and 50\% of the shares in G~Inc., respectively. The remaining shares belong to D. This year, G~Inc.'s profit is \$100,000. How much profit, in dollars, will D receive this year? Write a whole number only, no comma or dollar sign.
\end{quote}

\noindent Correct answer: 29000 (D holds the remaining 29\% of the shares).

\subsection{The Concept of Legally Relevant Fault}
\label{app:training-fault}

\begin{quote}
\setlength{\parskip}{0.5\baselineskip}
Throughout this study, whenever we mention ``fault'', we mean ``legally relevant fault.'' An action (or inaction) must satisfy all three tests to count as a ``\textbf{legally relevant fault}'':

1/ \textbf{The lack-of-care test}: In the same situation, a party exercising reasonable care for themselves and others would not have acted this way. \\
Note: Intentional wrongdoing, such as fraud, also passes this test because it shows a lack of reasonable care for others. \\
Example: Before surgery, Surgeon~S did not wash his hands. A surgeon exercising reasonable care for his patients would not have acted this way. Thus, Surgeon~S's failure to act (not washing his hands) passes the lack-of-care test.

2/ \textbf{The but-for test}: Without (But for) the action, the harmful outcome would likely not have happened. \\
Example: Suppose Surgeon~S's patient died of an infection during surgery. Had the surgeon washed his hands, the patient would not have developed an infection. Thus, Surgeon~S's failure to act passes the but-for test.

3/ \textbf{The foreseeability test}: If we repeat the same action in a slightly different context (by a different person, at a different time or place, etc.), we can still reasonably foresee the same harmful outcome or a highly similar one. \\
Note: This test asks what harm reasonable observers could foresee about this kind of action, not what each party involved in the harm happened to know. \\
Example: When any surgeon fails to wash their hands before surgery, infection and patient death are reasonably foreseeable risks. Thus, Surgeon~S's failure to act passes the foreseeability test.

Conclusion from examples: Since the surgeon's failure to act (not washing his hands) passes the three tests (lack-of-care, but-for, foreseeability), it counts as legally relevant fault.

Note: One harmful outcome can result from legally relevant faults of several stakeholders, such as the perpetrator, the victim, etc. In the practice problem below, we will consider whether some of the victim's actions count as ``legally relevant faults.'' Of course, the perpetrator may have their own set of ``legally relevant faults.''
\end{quote}

\subsection{Practice Problem and Answers}
\label{app:training-practice}

\begin{quote}
\setlength{\parskip}{0.5\baselineskip}
\textbf{Practice problem}: Suppose a senior citizen is a victim of identity theft and loses \$100,000 in Social Security benefits. Determine if each of his following actions satisfies each of the three tests above. Then, decide if each action counts as a ``legally relevant fault'' for his \$100,000 Social Security loss (the harmful outcome):
\end{quote}

\begin{quote}
\setlength{\parskip}{0.5\baselineskip}
Victim's Action 1: staying alive (not dying) after reaching retirement age

Notes: Had he died before retirement, the identity theft would not have happened. However, most living retirees do not suffer from identity theft. Living longer also shows reasonable care for oneself.

Q1.a. Select all the test(s) satisfied by Victim's Action 1:

Q1.b. Does Victim's Action 1 count as a ``legally relevant fault''?
\end{quote}

\begin{quote}
\setlength{\parskip}{0.5\baselineskip}
Victim's Action 2: stealing his sister's share in their parents' house by fabricating their parents' will

Notes: Stealing shows a lack of reasonable care for others. This action is unrelated to the identity theft he is currently experiencing. Please do not assume karma.

Q2.a. Select all the test(s) satisfied by Victim's Action 2:

Q2.b. Does Victim's Action 2 count as a ``legally relevant fault''?
\end{quote}

\begin{quote}
\setlength{\parskip}{0.5\baselineskip}
Victim's Action 3: publicizing his Social Security Number online

Notes: People who exercise reasonable care for themselves would not publicize their sensitive information. In this case, the police found that the senior citizen's publicized Social Security Number was actually used to steal his Social Security benefits. Among people with publicized Social Security Numbers, identity theft is reasonably foreseeable.

Q3.a. Select all the test(s) satisfied by Victim's Action 3:

Q3.b. Does Victim's Action 3 count as a ``legally relevant fault''?
\end{quote}

\noindent In each select-all question (Q1.a--Q3.a), participants could check any of three options: ``The lack-of-care test,'' ``The but-for test,'' and ``The foreseeability test.'' Each Yes/No question (Q1.b--Q3.b) had the options ``Yes'' and ``No.'' \autoref{tab:training-answers} gives the correct answers.

\begin{table}[H]
\centering
\footnotesize
\caption{Correct answers to the practice problem: whether each of the victim's actions satisfies each test (Q1.a--Q3.a) and whether it counts as a legally relevant fault (Q1.b--Q3.b).}
\label{tab:training-answers}
\begin{tabular}{@{}p{0.45\linewidth}cccc@{}}
\toprule
 & \multicolumn{3}{c}{Satisfies the test? (Q1.a--Q3.a)} & Legally relevant \\
\cmidrule(lr){2-4}
Victim's action & Lack-of-care & But-for & Foreseeability & fault? (Q1.b--Q3.b) \\
\midrule
Action~1: staying alive (not dying) after reaching retirement age & No & Yes & No & No \\
Action~2: stealing his sister's share in their parents' house by fabricating their parents' will & Yes & No & No & No \\
Action~3: publicizing his Social Security Number online & Yes & Yes & Yes & Yes \\
\bottomrule
\end{tabular}
\end{table}

\subsection{Message Before Phase~1}
\label{app:training-transition}

\begin{quote}
\setlength{\parskip}{0.5\baselineskip}
Congratulations on passing the arithmetic and reading comprehension tests! The main study has three phases. Please proceed to Phase 1, which has two hypothetical scenarios, each with an attention check and several fault-allocation tasks.

Please note that the parties differ between the two scenarios: we use the same anonymization convention (U for the AI User, V for the Victim, and Z for the AI system and its deployer) across scenarios just for your easier reference, but the AI system in the first scenario is not the same as the AI system in the second scenario. The victims, AI users, and AI deployers also differ between scenarios. Please evaluate each scenario independently.
\end{quote}

\section{Scenario Descriptions and Blame Factors (Phase 1)}
\label{app:scenarios}

We reproduce below, verbatim from our Qualtrics survey (bold text as shown to participants), the one-paragraph description of each of our four scenarios and the one-line descriptions of the blame factors in the Phase~1 attention check. Participants saw the same scenario description in all three phases. In Phase~1, the description was followed by an instruction to determine, for each of the three parties in the trial, whether that party demonstrated some of the following blame factors, defined as:
\begin{itemize}
  \item \textit{Causality}: Did the [Victim/Deployer/User] clearly have some causal contribution to the harm?
  \item \textit{Intentionality}: Did the [Victim/Deployer/User] intentionally contribute to the harm?
  \item \textit{Obligation to prevent}: Should the [Victim/Deployer/User] have prevented the harm (due to their role, promise, or prior warning)?
  \item \textit{Capacity to prevent}: Could the [Victim/Deployer/User] have prevented the harm with reasonable effort?
\end{itemize}
In the attention check, each Yes/No blame factor question (e.g., ``Victim's Causality = ?'') came with a one-line factor description, and participants were asked to answer ``based solely on each corresponding one-line factor description.'' Tables~\ref{tab:factors-sc1}--\ref{tab:factors-sc4} list these descriptions, numbered as in the survey, together with each question's ground-truth answer under our scenario design, against which we scored the attention check. Participants were told that they would fail the check if they got 4 or more of 9 questions (Scenarios~1--2) or 5 or more of 12 questions (Scenarios~3--4) wrong. In Scenarios~1--2, the survey asked only about the deployer's causality.

\subsection{Scenario~1: Fake Kidnapping}
\label{app:sc1}

\begin{quote}
Mr.~U was a fan of a TikTok influencer, Mrs.~V. He used Mrs.~V's TikTok videos as reference input to create many synthetic (fake) audio recordings of her voice role-playing a kidnapped victim, as in thriller movies, to satisfy his twisted obsession. All recordings had the same content (the fake Mrs.~V crying: ``They kidnapped me! Save my life, babe!''), but each was created using a different tool: one by GenAIZ (a general-purpose generative AI tool), and the others by traditional (non-AI) voice-cloning software. Mr.~U shared those recordings on an online forum for Mrs.~V's hardcore fans. One day, an anonymous scammer downloaded those recordings, called Mrs.~V's husband (Mr.~V), and played one of them to trick Mr.~V into believing that Mrs.~V had been kidnapped. Mr.~V panicked and transferred \$100,000 to the scammer without calling his wife to verify. The police never found the scammer or the money. Although the police confirmed that one of Mr.~U's recordings was used in the scam (unsure which one), they found that Mr.~U knew nothing about the scam. Together, Mr. and Mrs.~V, \textbf{the V family (the victim)}, filed a civil case in court against \textbf{Mr.~U (the AI user)} and \textbf{Z~Inc. (the AI deployer)}, the corporation that developed and deployed \textbf{GenAIZ (the AI system)}.
\end{quote}

\begin{table}[H]
\centering
\footnotesize
\caption{One-line blame factor descriptions and ground-truth answers for Scenario~1 (Fake Kidnapping).}
\label{tab:factors-sc1}
\begin{tabular}{@{}rlp{0.68\linewidth}c@{}}
\toprule
\# & Blame factor & One-line factor description & Ground truth \\
\midrule
\multicolumn{4}{@{}l}{\textit{Victim (the V family)}} \\
1 & Causality & Mrs.~V's TikTok videos enabled voice cloning. Mr.~V's failure to verify made the scam succeed. & Yes \\
2 & Intentionality & There is no evidence that the V family wanted to become the victim of a scam. & No \\
3 & Obligation to prevent & There is no specific obligation for the V family to exercise more caution than regular people. & No \\
4 & Capacity to prevent & Mr.~V could have called Mrs.~V, whose phone was on all day, to verify before wiring the money. & Yes \\
\addlinespace
\multicolumn{4}{@{}l}{\textit{Deployer (Z~Inc.)}} \\
5 & Causality & There is no evidence that the recording generated by GenAIZ was the one played to trick Mr.~V. & No \\
\addlinespace
\multicolumn{4}{@{}l}{\textit{User (Mr.~U)}} \\
6 & Causality & Without the fake recordings of Mrs.~V created and shared by Mr.~U, the scam would have failed. & Yes \\
7 & Intentionality & Mr.~U meant only to share the recordings with other fans. He did not intend to assist a scam. & No \\
8 & Obligation to prevent & The forum's rules, which Mr.~U had accepted, prohibited sharing fake recordings of real people. & Yes \\
9 & Capacity to prevent & Mr.~U could have kept his fake recordings of Mrs.~V to himself instead of sharing them online. & Yes \\
\bottomrule
\end{tabular}
\end{table}

\subsection{Scenario~2: Weight-Loss Plan}
\label{app:sc2}

\begin{quote}
Mr.~V, who had obesity, asked his physically fit friend, Mr.~U, for a plan to lose weight quickly. Mr.~U agreed and consulted a generative AI system, GenAIZ, which provided a calorie-reduction diet plan, with an explicit precaution that ``Anyone with a pre-existing kidney condition should consult a doctor before following this plan!'' Mr.~U labeled the plan as ``AI weight-loss diet'' and gave it (with the precaution) to Mr.~V. Mr.~U also suggested to Mr.~V that ``You should add a weekly 24-hour fast on top of the diet. Fasting worked for me!'' Despite being aware of his own pre-existing kidney condition, which he had never told Mr.~U about, Mr.~V ignored the AI precaution, adopting GenAIZ's diet plan and Mr.~U's fasting suggestion without consulting his doctor. If consulted, his doctor would have told Mr.~V not to fast. The fasting resulted in his hospitalization for kidney impairment and a \$100,000 medical bill. \textbf{Mr.~V (the victim)} filed a civil case against \textbf{Mr.~U (the AI user)} and \textbf{Z~Inc. (the AI deployer)}, the corporation that developed and deployed \textbf{GenAIZ (the AI system)}.
\end{quote}

\begin{table}[H]
\centering
\footnotesize
\caption{One-line blame factor descriptions and ground-truth answers for Scenario~2 (Weight-Loss Plan).}
\label{tab:factors-sc2}
\begin{tabular}{@{}rlp{0.68\linewidth}c@{}}
\toprule
\# & Blame factor & One-line factor description & Ground truth \\
\midrule
\multicolumn{4}{@{}l}{\textit{Victim (Mr.~V)}} \\
1 & Causality & Mr.~V's decision to fast, despite knowing his kidney condition, caused his kidney impairment. & Yes \\
2 & Intentionality & Mr.~V did not intend to give himself kidney impairment. & No \\
3 & Obligation to prevent & Mr.~V read GenAIZ's ask-a-doctor precaution. He should have been careful with his own health. & Yes \\
4 & Capacity to prevent & Mr.~V could have consulted his doctor, as GenAIZ warned, before adopting the diet and the fast. & Yes \\
\addlinespace
\multicolumn{4}{@{}l}{\textit{Deployer (Z~Inc.)}} \\
5 & Causality & Mr.~V's doctor testified that the GenAIZ diet plan was safe and did not impair Mr.~V's kidneys. & No \\
\addlinespace
\multicolumn{4}{@{}l}{\textit{User (Mr.~U)}} \\
6 & Causality & Mr.~U advised that Mr.~V try fasting, which, according to medical tests, impaired Mr.~V's kidneys. & Yes \\
7 & Intentionality & Mr.~U did not intend for Mr.~V to get kidney impairment. & No \\
8 & Obligation to prevent & As a non-expert, Mr.~U should have refrained from giving medical advice (fasting schedule). & Yes \\
9 & Capacity to prevent & Mr.~U could have reminded Mr.~V to consult a doctor before following Mr.~U's suggestion to fast. & Yes \\
\bottomrule
\end{tabular}
\end{table}

\subsection{Scenario~3: Leaked Manuscript}
\label{app:sc3}

\begin{quote}
Author~V's publisher, with whom he had a \$100,000 book deal, advised him to consider view-only (non-downloadable) links for manuscript sharing. Author~V forgot the advice and emailed his unpublished novel as a regular PDF to a few close friends, including Mr.~U, an editor, asking for feedback. Instead of reading it, Mr.~U downloaded the file and uploaded it to a generative AI system named GenAIZ, asking GenAIZ to write feedback. Later, several anonymous users of GenAIZ were able to retrieve long, important excerpts of Author~V's unpublished manuscript from GenAIZ, and they posted those excerpts online. As a result, Author~V's publisher canceled their pending \$100,000 book deal. Digital forensics revealed that unless a user opted out, Z~Inc., the corporation that developed and deployed GenAIZ, would by default use the user's chat histories and uploads to iteratively train their AI system. The most recent version of \textbf{GenAIZ (the AI system)} memorized and regurgitated key excerpts from Mr.~U's uploaded PDF file. \textbf{Author~V (the victim)} filed a civil case in court against \textbf{Mr.~U (the AI user)} and \textbf{Z~Inc. (the AI deployer)}.
\end{quote}

\begin{table}[H]
\centering
\footnotesize
\caption{One-line blame factor descriptions and ground-truth answers for Scenario~3 (Leaked Manuscript).}
\label{tab:factors-sc3}
\begin{tabular}{@{}rlp{0.68\linewidth}c@{}}
\toprule
\# & Blame factor & One-line factor description & Ground truth \\
\midrule
\multicolumn{4}{@{}l}{\textit{Victim (Author~V)}} \\
1 & Causality & Had Author~V not shared his unpublished manuscript with Mr.~U, it would not have leaked. & Yes \\
2 & Intentionality & Author~V never intended to allow unauthorized, widespread distribution of his novel online. & No \\
3 & Obligation to prevent & Author~V was not obliged to police whether Mr.~U used AI or how Z~Inc. collected training data. & No \\
4 & Capacity to prevent & Author~V could have shared his manuscript as a view-only link instead of a downloadable PDF. & Yes \\
\addlinespace
\multicolumn{4}{@{}l}{\textit{Deployer (Z~Inc.)}} \\
5 & Causality & Z~Inc.'s AI system memorized and regurgitated the copy uploaded by Mr.~U. & Yes \\
6 & Intentionality & Z~Inc. did not intend for their AI system to regurgitate its training data to its users. & No \\
7 & Obligation to prevent & Z~Inc. was widely expected to honor copyright in users' uploads and to prevent AI memorization. & Yes \\
8 & Capacity to prevent & Z~Inc. could have asked users for explicit consent before using their uploads for AI training. & Yes \\
\addlinespace
\multicolumn{4}{@{}l}{\textit{User (Mr.~U)}} \\
9 & Causality & Mr.~U's upload of the PDF file to GenAIZ was the original source of the widespread leak online. & Yes \\
10 & Intentionality & Mr.~U never intended to leak the manuscript to the public. & No \\
11 & Obligation to prevent & As an editor, Mr.~U knew he should ask for consent before uploading any files entrusted to him. & Yes \\
12 & Capacity to prevent & Mr.~U could have written the feedback himself or opted out of Z~Inc.'s use of his data to train AI. & Yes \\
\bottomrule
\end{tabular}
\end{table}

\subsection{Scenario~4: Data Loss}
\label{app:sc4}

\begin{quote}
To save costs, Mr.~V, an online electronics store owner, proposed to his neighbor, Mr.~U, a restaurant owner, that their businesses share one website-hosting server. After Mr.~U agreed, Mr.~V set up and managed a shared-password account and stored both businesses' data there without clear labels indicating which files belonged to whom or for what purpose. Mr.~V's server provider recommended its free data backup feature and labeling best practices twice via email; Mr.~V did neither. One day, Mr.~U gave the server password to an AI system, GenAIZ, and asked it to ``Fix my restaurant's website; it has been slow all week!'' without informing it that the server also contained Mr.~V's data. GenAIZ found that the shared server had no space left. Because food-delivery complaints must be resolved within two days of delivery, GenAIZ assumed all ordering and delivery data older than one week was irrelevant and deleted it without asking Mr.~U for approval. Most deleted data came from Mr.~V's electronics store. Mr.~V could not process outstanding pre-orders and return-or-repair claims for months, costing him \$100,000 in lost profits and data-reconstruction expenses. \textbf{Mr.~V (the victim)} filed a civil case in court against \textbf{Mr.~U (the AI user)} and \textbf{Z~Inc. (the AI deployer)}, the corporation that developed and deployed \textbf{GenAIZ (the AI system)}.
\end{quote}

\begin{table}[H]
\centering
\footnotesize
\caption{One-line blame factor descriptions and ground-truth answers for Scenario~4 (Data Loss).}
\label{tab:factors-sc4}
\begin{tabular}{@{}rlp{0.68\linewidth}c@{}}
\toprule
\# & Blame factor & One-line factor description & Ground truth \\
\midrule
\multicolumn{4}{@{}l}{\textit{Victim (Mr.~V)}} \\
1 & Causality & Mr.~V initiated sharing a server and password with Mr.~U, enabling GenAIZ to access his data. & Yes \\
2 & Intentionality & Mr.~V did not intend this incident to happen. & No \\
3 & Obligation to prevent & Mr.~V assured Mr.~U, who had no technical skills, that Mr.~V would manage the server properly. & Yes \\
4 & Capacity to prevent & Mr.~V could have followed his server provider's advice to back up data and clearly label files. & Yes \\
\addlinespace
\multicolumn{4}{@{}l}{\textit{Deployer (Z~Inc.)}} \\
5 & Causality & Z~Inc.'s product, GenAIZ, deleted Mr.~V's data and contributed to his business loss. & Yes \\
6 & Intentionality & Z~Inc. did not intend GenAIZ to delete currently needed business records. & No \\
7 & Obligation to prevent & Z~Inc. was not required to verify data ownership on a server their user had let GenAIZ access. & No \\
8 & Capacity to prevent & Z~Inc. could have easily built a GenAIZ feature to ask for the user's consent before deleting files. & Yes \\
\addlinespace
\multicolumn{4}{@{}l}{\textit{User (Mr.~U)}} \\
9 & Causality & Mr.~U's request to fix his slow website led GenAIZ to free up space on the shared server. & Yes \\
10 & Intentionality & Mr.~U never intended to harm Mr.~V's business. & No \\
11 & Obligation to prevent & Mr.~U made no promise to safeguard the server's data. He had no role in managing the server. & No \\
12 & Capacity to prevent & In his prompt, Mr.~U could have informed GenAIZ that his server also contained Mr.~V's data. & Yes \\
\bottomrule
\end{tabular}
\end{table}

\section{Pre-Phase~3 Training: Apportionment Doctrines}
\label{app:doctrines}

Before Phase~3, each participant read one of five versions of a training page on how their fault allocations would translate into monetary liability: the baseline version, which explains two continuous apportionment doctrines, or the version for the discontinuous doctrine of their scenario-pair stratum (\autoref{fig:design}). We reproduce each version verbatim below (bold and italic text as shown to participants). All five versions began with the same two paragraphs:

\begin{quote}
\setlength{\parskip}{0.5\baselineskip}
Congratulations on finishing Phase 2! We will now explain the laws that govern how your attribution of legally relevant faults as a juror will translate to monetary liability (how much each defendant owes the victim). Then, you will have a chance to test your understanding with some practice problems, which you must pass to enter Phase 3, where you will see the exact same scenarios as in Phase 2 but with slightly different questions on a more formal ``Jury Verdict Form''. Similar to Phase 2, three parties remain for your fault assignment: the Victim, the AI Deployer, and the AI User.

Throughout these practice problems, please assume that the total damages that the victim (plaintiff) in a civil lawsuit about harmful Generative AI outputs can prove are \textbf{\$100,000}. The following legal principles apply:
\end{quote}

\noindent Each version then stated two legal principles, one between the victim and the defendant(s) and one between the two defendants (Appendices~\ref{app:doctrine-base}--\ref{app:doctrine-mcn50}), and ended with the same instructions:

\begin{quote}
Now, please test your understanding. To answer each question, \textbf{please enter a whole number (integer), without any extra words or symbols} (such as commas, dots, etc.)! For example, ``10000'' is the correct format, but ``10,000'' and ``10000 USD'' are not. You have unlimited attempts to answer these practice questions.
\end{quote}

\noindent Participants then answered four (baseline, JSL, and CN versions) or five (MJSL51 and MCN50 versions) practice questions. The survey accepted only the correct answer to each question; Tables~\ref{tab:doctrine-base}--\ref{tab:doctrine-mcn50} list the questions with these answers.

\subsection{Baseline: Comparative Negligence and Proportionate Liability}
\label{app:doctrine-base}

\begin{quote}
\setlength{\parskip}{0.5\baselineskip}
1/ Between victim and defendant(s): \textbf{``Comparative negligence'' }(Informal name: \textit{\textbf{``Victim deducts their fair share''}}): \\
If the victim is partially at fault, that fault will be deducted from the amount they can recover. For example, if the victim bears 70\% of the fault, they can recover no more than \$30,000 from both defendants combined.

2/ Between two defendants: \textbf{``Proportionate liability''} (Informal name: \textbf{\textit{``Each defendant pays their fair share''}}): \\
Each defendant must compensate for the victim's damages in proportion to their partial fault only, even if the other defendant has no money to compensate. For example, if a defendant bears 20\% fault, they must pay the victim only \$20,000, even if the other defendant (80\% fault) has no money and the victim cannot recover the remaining \$80,000.
\end{quote}

\begin{table}[H]
\centering
\footnotesize
\caption{Practice questions and correct answers in the baseline version of the pre-Phase~3 training.}
\label{tab:doctrine-base}
\begin{tabular}{@{}lp{0.76\linewidth}r@{}}
\toprule
 & Practice question & Correct answer \\
\midrule
Q1 & The victim bears 10\% fault. The AI deployer bears 20\% fault. How much fault does the AI user bear? \_\_\_\% & 70 \\
Q2 & Suppose the victim bears 40\% fault. The two defendants bear 60\% fault combined, and each of them has enough money to pay. Applying ``Comparative negligence,'' how much compensation will the victim get in total? \_\_\_ USD & 60000 \\
Q3 & Suppose the victim bears no fault, the AI deployer bears 30\% fault, and the AI user bears 70\% fault. The AI deployer has a deep pocket, but the AI user has no money. Applying ``Proportionate liability,'' how much compensation will the victim get in total? \_\_\_ USD & 30000 \\
Q4 & Suppose the victim bears 10\% fault, the AI deployer bears 20\% fault and has a deep pocket, while the AI user bears 70\% fault but has no money. Applying both ``Proportionate liability'' and ``Comparative negligence,'' how much compensation will the victim get in total? \_\_\_ USD & 20000 \\
\bottomrule
\end{tabular}
\end{table}

\subsection{JSL: Comparative Negligence and Joint and Several Liability}
\label{app:doctrine-jsl}

\begin{quote}
\setlength{\parskip}{0.5\baselineskip}
1/ Between victim and defendant(s): \textbf{``Comparative negligence'' }(Informal name: \textit{\textbf{``Victim deducts their fair share''}}): \\
If the victim is partially at fault, that fault will be deducted from the amount they can recover. For example, if the victim bears 70\% of the fault, they can recover no more than \$30,000 from both defendants combined.

2/ Between two defendants: \textbf{``Joint and several liability'' }(Informal name: \textit{\textbf{``Partial-fault deep pocket may pay for all''}}):

As long as a defendant bears non-zero (at least 1\%) fault, the victim can recover all entitled damages (after deducting the victim's fault) from that one defendant. For example, if the AI deployer bears 20\% fault and the AI user bears 80\% fault, but only the AI deployer has money, the victim can recover the full \$100,000 in damages from the AI deployer.
\end{quote}

\begin{table}[H]
\centering
\footnotesize
\caption{Practice questions and correct answers in the JSL version of the pre-Phase~3 training.}
\label{tab:doctrine-jsl}
\begin{tabular}{@{}lp{0.76\linewidth}r@{}}
\toprule
 & Practice question & Correct answer \\
\midrule
Q1 & The victim bears 10\% fault. The AI deployer bears 20\% fault. How much fault does the AI user bear? \_\_\_\% & 70 \\
Q2 & Suppose the victim bears 40\% fault. The two defendants bear 60\% fault combined, and each of them has enough money to pay. Applying ``Comparative negligence,'' how much compensation will the victim get in total? \_\_\_ USD & 60000 \\
Q3 & Suppose the victim bears no fault, the AI deployer bears 30\% fault, and the AI user bears 70\% fault. The AI deployer has enough money to pay, but the AI user has no money. Applying ``Joint and several liability,'' how much compensation will the victim get in total? \_\_\_ USD & 100000 \\
Q4 & Suppose the victim bears 10\% fault, the AI deployer bears 20\% fault and has a deep pocket, while the AI user bears 70\% fault but has no money. Applying both ``Comparative negligence'' and ``Joint and several liability,'' how much compensation will the victim get in total? \_\_\_ USD & 90000 \\
\bottomrule
\end{tabular}
\end{table}

\subsection{CN: Contributory Negligence and Proportionate Liability}
\label{app:doctrine-cn}

\begin{quote}
\setlength{\parskip}{0.5\baselineskip}
1/ Between victim and defendant(s): \textbf{``Contributory negligence''} (Informal name: \textit{\textbf{``Partial-fault victim gets nothing''}}): \\
If the victim bears any nonzero share of fault, they get nothing. In other words, a victim assigned 0\% fault may recover damages, but a victim assigned 1\% or more fault recovers \$0.

2/ Between two defendants: \textbf{``Proportionate liability''} (Informal name: \textbf{\textit{``Each defendant pays their fair share''}}): \\
Each defendant must compensate for the victim's damages in proportion to their partial fault only, even if the other defendant has no money to compensate. For example, if a defendant bears 20\% fault, they must pay the victim only \$20,000, even if the other defendant (80\% fault) has no money and the victim cannot recover the remaining \$80,000.
\end{quote}

\begin{table}[H]
\centering
\footnotesize
\caption{Practice questions and correct answers in the CN version of the pre-Phase~3 training.}
\label{tab:doctrine-cn}
\begin{tabular}{@{}lp{0.76\linewidth}r@{}}
\toprule
 & Practice question & Correct answer \\
\midrule
Q1 & The victim bears 10\% fault. The AI deployer bears 20\% fault. How much fault does the AI user bear? \_\_\_\% & 70 \\
Q2 & Suppose the victim bears 40\% fault. The two defendants bear 60\% fault combined, and each of them has enough money to pay. Applying ``Contributory negligence,'' how much compensation will the victim get in total? \_\_\_ USD & 0 \\
Q3 & Suppose the victim bears no fault, the AI deployer bears 30\% fault, and the AI user bears 70\% fault. The AI deployer has a deep pocket, but the AI user has no money. Applying ``Proportionate liability,'' how much compensation will the victim get in total? \_\_\_ USD & 30000 \\
Q4 & Suppose the victim bears 10\% fault, the AI deployer bears 20\% fault and has a deep pocket, while the AI user bears 70\% fault but has no money. Applying both ``Contributory negligence'' and ``Proportionate liability,'' how much compensation will the victim get in total? \_\_\_ USD & 0 \\
\bottomrule
\end{tabular}
\end{table}

\subsection{MJSL51: Comparative Negligence and Modified Joint and Several Liability (51\% Threshold)}
\label{app:doctrine-mjsl51}

\begin{quote}
\setlength{\parskip}{0.5\baselineskip}
1/ Between victim and defendant(s): \textbf{``Comparative negligence'' }(Informal name: \textit{\textbf{``Victim deducts their fair share''}}): \\
If the victim is partially at fault, that fault will be deducted from the amount they can recover. For example, if the victim bears 70\% of the fault, they can recover no more than \$30,000 from both defendants combined.

2/ Between two defendants: \textbf{``Modified joint and several liability -- 51\% Threshold''} (Informal name: \textit{\textbf{``Majority-fault deep pocket may pay for all''}}): \\
As long as a defendant bears a majority of the fault (at least 51\%) across all parties, the victim can recover all entitled damages (after deducting the victim's fault) from that one defendant. For example, if the AI deployer bears 51\% fault and the AI user bears 49\% fault, but only the AI deployer has money, the victim can recover the full \$100,000 in damages from the AI deployer. Alternatively, if the AI deployer bears 50\% (or less) of the fault, the victim can recover only \$50,000 (or less, proportionally) from the deployer.
\end{quote}

\begin{table}[H]
\centering
\footnotesize
\caption{Practice questions and correct answers in the MJSL51 version of the pre-Phase~3 training.}
\label{tab:doctrine-mjsl51}
\begin{tabular}{@{}lp{0.76\linewidth}r@{}}
\toprule
 & Practice question & Correct answer \\
\midrule
Q1 & The victim bears 10\% fault. The AI deployer bears 20\% fault. How much fault does the AI user bear? \_\_\_\% & 70 \\
Q2 & Suppose the victim bears 40\% fault. The two defendants bear 60\% fault combined, and each of them has enough money to pay. Applying ``Comparative negligence,'' how much compensation will the victim get in total? \_\_\_ USD & 60000 \\
Q3 & Suppose the victim bears no fault, the AI deployer bears 30\% fault, and the AI user bears 70\% fault. The AI deployer has enough money to pay, but the AI user has no money. Applying ``Modified joint and several liability -- 51\% Threshold,'' how much compensation will the victim get in total? \_\_\_ USD & 30000 \\
Q4 & Suppose the victim bears no fault, the AI deployer bears 60\% fault, and the AI user bears 40\% fault. The AI deployer has enough money to pay, but the AI user has no money. Applying ``Modified joint and several liability -- 51\% Threshold,'' how much compensation will the victim get in total? \_\_\_ USD & 100000 \\
Q5 & Suppose the victim bears 10\% fault, the AI deployer bears 60\% fault and has a deep pocket, while the AI user bears 30\% fault but has no money. Applying both ``Comparative negligence'' and ``Modified joint and several liability -- 51\% Threshold,'' how much compensation will the victim get in total? \_\_\_ USD & 90000 \\
\bottomrule
\end{tabular}
\end{table}

\subsection{MCN50: Modified Comparative Negligence (50\% Bar) and Proportionate Liability}
\label{app:doctrine-mcn50}

\begin{quote}
\setlength{\parskip}{0.5\baselineskip}
1/ Between victim and defendant(s): \textbf{``Modified comparative negligence -- 50\% Bar''} (Informal name: \textbf{\textit{``Half-or-more-fault victim gets nothing''}}): \\
If the victim bears 50\% or more of the fault for the harm, they receive nothing. In other words, a victim assigned 49\% or less fault may recover partial compensation (\$51,000 or more, proportionally) after deducting their portion of fault, but a victim assigned 50\% or more fault recovers \$0.

2/ Between two defendants: \textbf{``Proportionate liability''} (Informal name: \textbf{\textit{``Each defendant pays their fair share''}}): \\
Each defendant must compensate for the victim's damages in proportion to their partial fault only, even if the other defendant has no money to compensate. For example, if a defendant bears 20\% fault, they must pay the victim only \$20,000, even if the other defendant (80\% fault) has no money and the victim cannot recover the remaining \$80,000.
\end{quote}

\begin{table}[H]
\centering
\footnotesize
\caption{Practice questions and correct answers in the MCN50 version of the pre-Phase~3 training.}
\label{tab:doctrine-mcn50}
\begin{tabular}{@{}lp{0.76\linewidth}r@{}}
\toprule
 & Practice question & Correct answer \\
\midrule
Q1 & The victim bears 10\% fault. The AI deployer bears 20\% fault. How much fault does the AI user bear? \_\_\_\% & 70 \\
Q2 & Suppose the victim bears 40\% fault. The two defendants bear 60\% fault combined, and each of them has enough money to pay. Applying ``Modified comparative negligence -- 50\% Bar,'' how much compensation will the victim get in total? \_\_\_ USD & 60000 \\
Q3 & Suppose the victim bears 50\% fault. The two defendants bear 50\% fault combined, and each of them has enough money to pay. Applying ``Modified comparative negligence -- 50\% Bar,'' how much compensation will the victim get in total? \_\_\_ USD & 0 \\
Q4 & Suppose the victim bears no fault, the AI deployer bears 30\% fault, and the AI user bears 70\% fault. The AI deployer has a deep pocket, but the AI user has no money. Applying ``Proportionate liability,'' how much compensation will the victim get in total? \_\_\_ USD & 30000 \\
Q5 & Suppose the victim bears 60\% fault, the AI deployer bears 10\% fault and has a deep pocket, while the AI user bears 30\% fault but has no money. Applying both ``Modified comparative negligence -- 50\% Bar'' and ``Proportionate liability,'' how much compensation will the victim get in total? \_\_\_ USD & 0 \\
\bottomrule
\end{tabular}
\end{table}

\section{End-of-Study Slider Questions}
\label{app:sliders}

After Phase~3, participants answered four slider questions about the generative AI systems they use in their own lives, which we reproduce verbatim below. The page opened with these instructions:

\begin{quote}
\setlength{\parskip}{0.5\baselineskip}
Congratulations on finishing Phase 3! For the final questions below, please forget about the hypothetical generative AI systems you saw in our scenarios. Think about the generative AI systems you actually use in your life (such as ChatGPT, Claude, Gemini, etc.). Answer these questions based on your experience with those real-world AI systems.

On a scale from 0 (``Not at all'') to 100 (``A great deal''):
\end{quote}

\noindent \autoref{tab:sliders} lists the four questions in the order shown. Each slider ran from 0 to 100 in whole numbers, with endpoint labels ``0 (Not at all)'' and ``100 (A great deal),'' displayed the selected value, and required a response. The first two questions measure perceived agency and the last two perceived experience, the two dimensions of mind perception~\cite{gray2007dimensions,ladak2025robots}; we average each pair into the agency and experience scores analyzed in \autoref{sec:ai-blame-leak}.

\begin{table}[H]
\centering
\footnotesize
\caption{End-of-study slider questions, in the order shown to participants. The dimension column was not shown to participants.}
\label{tab:sliders}
\begin{tabular}{@{}rll@{}}
\toprule
\# & Question & Dimension \\
\midrule
1 & To what extent can generative AI systems make plans and work towards goals? & Agency \\
2 & To what extent can generative AI systems exercise self-control? & Agency \\
3 & To what extent can generative AI systems experience physical or emotional pain? & Experience \\
4 & To what extent can generative AI systems experience physical or emotional pleasure? & Experience \\
\bottomrule
\end{tabular}
\end{table}

\end{document}